\documentclass{aa}  
\usepackage{natbib,epsfig,graphicx,color,psfrag,pdflscape}
\usepackage{amsmath}
\usepackage[varg]{txfonts}
\bibpunct{(}{)}{,}{a}{}{,}

\definecolor{cerulean}{rgb}{0.11,0.67,0.84}
\definecolor{bluegray}{rgb}{0.4,0.6,0.8}
\definecolor{forest}{rgb}{0.0,0.26,0.15}
\definecolor{highlight}{rgb}{0.87,1.0,0.0}
\definecolor{darkblue}{rgb}{0.0,0.2,0.6}

\newcommand{\rsun}{\ensuremath{R_{\odot}}}

\newcommand{\Teff}{\ensuremath{T_{\rm eff}}}
\newcommand{\vinf}{\ensuremath{v_{\infty}}}
\newcommand{\mdot}{\ensuremath{\dot{M}}}

\newcommand{\mdu}{\ensuremath{10^{-6}\,M_{\odot} {\rm yr}^{-1}}}

\newcommand{\beq}{\begin{equation}}
\newcommand{\eeq}{\end{equation}}
\newcommand{\beqa}{\begin{eqnarray}}
\newcommand{\eeqa}{\end{eqnarray}}

\newcommand{\nbeq}{\begin{equation*}}
\newcommand{\neeq}{\end{equation*}}

\newcommand{\kms}{\ensuremath{{\rm km}\,{\rm s}^{-1}}}
\newcommand{\cms}{\ensuremath{{\rm cm}\,{\rm s}^{-2}}}

\newcommand{\III}{\,{\sc iii}}
\newcommand{\IV}{\,{\sc iv}}

\newcommand{\cions}{C\,{\sc ii/iii/iv}}
\newcommand{\cionss}{C\,{\sc ii/iii/iv/v}}

\newcommand{\HeI} {He\,{\sc i}}
\newcommand{\HeII}{He\,{\sc ii}}

\newcommand{\AlII}{Al\,{\sc ii}}
\newcommand{\AlIII}{Al\,{\sc iii}}

\newcommand{\ArII}{Ar\,{\sc ii}}

\newcommand{\CII}{C\,{\sc ii}}
\newcommand{\CIII}{C\,{\sc iii}}
\newcommand{\CIV}{C\,{\sc iv}}
\newcommand{\CV}{C\,{\sc v}}

\newcommand{\NII}{N\,{\sc ii}}
\newcommand{\NIII}{N\,{\sc iii}}

\newcommand{\NV}{N\,{\sc v}}

\newcommand{\OII}{O\,{\sc ii}}
\newcommand{\OIII}{O\,{\sc iii}}

\newcommand{\NeIII}{Ne\,{\sc iii}}

\newcommand{\SiIII}{Si\,{\sc iii}}

\newcommand{\SII}{S\,{\sc ii}}
\newcommand{\SIII}{S\,{\sc iii}}

\newcommand{\SVI}{S\,{\sc vi}}

\newcommand{\BaII}{Ba\,{\sc ii}}

\newcommand{\TiI}{Ti\,{\sc i}}

\newcommand{\LiII}{Liii\,{\sc ii}}

\newcommand{\MnI}{Mn\,{\sc i}}
\newcommand{\MnII}{Mn\,{\sc ii}}

\newcommand{\ScIII}{Sc\,{\sc iii}}

\newcommand{\FeI}{Fe\,{\sc i}}
\newcommand{\FeII}{Fe\,{\sc ii}}

\newcommand{\Ha} {H$_{\rm \alpha}$}

\newcommand{\Rstar}{\ensuremath{R_{\ast}}}

\newcommand{\Rmin}{\ensuremath{R_{\rm min}}}

\newcommand{\logg}{\ensuremath{\log g}}

\newcommand{\Tshockmax}{\ensuremath{T_{\rm s}^{\infty}}}

\newcommand{\vesc}{\ensuremath{v_{\rm esc}}}

\newcommand{\vsini}{\ensuremath{v{\thinspace}\sin{\thinspace}i}}
\newcommand{\vturb}{\ensuremath{v_{\rm turb}}}

\newcommand{\vmac}{\ensuremath{v_{\rm mac}}}

\newcommand{\fcl}{\ensuremath{f_{\rm cl}}}

\newcommand{\taur}{\ensuremath{\tau_{\rm Ross}}}

\def\apjl{ApJ}%
\def\apjs{ApJS}%
\def\ao{Appl.~Opt.}%
\def\aap{A\&A}%
\begin{document}
%
\title{Carbon line formation and spectroscopy in O-type stars}

\author{Luiz P. Carneiro, J. Puls, \and T. L. Hoffmann}

\institute{Universit\"atssternwarte M\"unchen, Scheinerstr. 1, 81679 M\"unchen, 
           Germany, \email{luiz@usm.uni-muenchen.de}}

\date{}

\abstract 
{The determination of chemical abundances constitutes a fundamental
requirement for obtaining a complete picture of a star. 
Particularly in massive stars, CNO abundances are of prime
interest, due to the nuclear CNO-cycle and various mixing processes
which bring these elements to the surface. The
precise determination of carbon abundances, together with N and O, is
thus a key ingredient to understanding different phases of stellar
evolution.}
{We aim at enabling a reliable carbon spectroscopy for our unified
NLTE atmosphere code {\sc FASTWIND}.
}
{We develop a new carbon model atom including \cionss, and discuss
specific problems related to carbon spectroscopy in O-type stars. We
describe different tests performed to examine the reliability of our
implementation, and investigate which mechanisms influence the carbon
ionization balance. By comparing with high-resolution spectra from six
O-type stars, we check in how far observational constraints can be
reproduced by our new carbon line synthesis.
}
{Carbon lines are even more sensitive to a variation of \Teff, \logg,
and \mdot, than hydrogen/helium lines.  We are able to reproduce most
of the observed lines from our stellar sample, and to estimate those 
specific carbon abundances which bring the lines from different ions
into agreement (three stages in parallel for cool objects, two for intermediate
O-types). For hot dwarfs and supergiants earlier than
O7, X-rays from wind-embedded shocks can have an impact on the
synthesized line strengths, particularly for \CIV, potentially
affecting the abundance determination. Dielectronic recombination has
a significant impact on the ionization balance in the wind. }
{We have demonstrated our capability to derive realistic carbon
abundances by means of {\sc FASTWIND}, using our recently developed
model atom. We found that complex effects can have a strong influence
on the carbon ionization balance in hot stars. For a further
understanding, the UV range needs to be explored as well.  
By means of detailed nitrogen and oxygen model atoms available to
use, we will be able to perform a complete CNO abundance analysis
for larger samples of massive stars, and to provide constraints on
corresponding evolutionary models and aspects.}


\titlerunning{Carbon spectroscopy of O-stars}
\authorrunning{Carneiro et al.}

\maketitle
%
%
\section{Introduction} 
\label{Introduction}

Quantitative spectroscopy provides decisive constraints on our
understanding of stellar evolution, chemical composition, and
nucleosynthesis. The analysis of stellar spectra using atmospheric
models tests the accuracy of present theoretical knowledge in this
regard. Therefore, any further theoretical development relies, to a
significant part, on the accuracy of data which describe the atomic
processes present in a thermodynamic system. Any inconsistency or
imprecision of the data directly affects a realistic representation of
nature. 

However, the calculation of atmospheric models is complex. In our
working field -- hot stars -- the strong radiation field leads to
non-LTE effects and causes a radiation-driven wind. This situation
can be handled by different codes, as for example {\sc CMFGEN}
\citep{hilliermiller98}, {\sc PHOENIX} \citep{haus92}, {\sc PoWR}
\citep{Graf02}, WM-{\sc basic} \citep{pauldrach01}, and {\sc FASTWIND}
\citep{puls05, rivero12}.  A brief comparison of these different codes
is given by \citet{puls09}.
 
Precise spectroscopic analysis (by means of accurate atmospheric
models) can lead to important conclusions about the chemical
composition of galaxies \citep{chiappini01,chiappini02} and 
corresponding metallicity gradients \citep{daflon04b}, especially when
performed using observations of early-type stars. Furthermore, it can
also give insights into mixing processes. At least in single stars,
the surface chemical composition is controlled by the efficiency of
mixing processes, which to a large part are associated with stellar
rotation. A high rotational velocity favors the transport of metals
from the stellar core to the surface, and consequentially the chemical
enrichment of the photosphere \citep{MaederVI,MeynetMaeder00}. 

In massive stars, nitrogen is a decisive indicator of such enrichment.
\citet{rivero11} investigated the formation of \NIII\ 4634-4640-4642,
and derived nitrogen abundances of O stars in the Magellanic Clouds
with a set of \NII-\III-\IV\ lines. However, even more precise
constraints on stellar evolution can be obtained using the N/C ratio,
since it is less sensitive to the initial metal content, compared to
N/H \citep{martins12}. Especially the combination of N/C vs. N/O (see
\citealt{przybilla10, Maeder14}) gives strong constraints on the
enrichment and mixing history of CNO material \citep{martins15a}, and
allows individual spectroscopic abundances to be tested. 

For these (and other) objectives, we have developed a new carbon model
atom to be used in spectroscopic analysis by means of {\sc FASTWIND},
suitable for the early B- and the complete O-star regime. 

Carbon plays a special role within the light elements. It is the basis
of all organic chemistry, but it is also essential for the
nucleosynthesis of H into He through the CNO cycle in massive hot
stars. Unfortunately, however, the analysis of carbon in such stars is
complicated, mainly because the number of carbon lines detectable in
O-type spectra is even smaller to the number of nitrogen lines. 

\citet{unsold42} pioneered the analysis of carbon spectra from 
early-type stars. Since then, numerous studies aimed at the same
objective, and we highlight here the contribution by \citet{nieva08},
which is the last in a series of three publications dedicated to
developing and applying a carbon model atom within the spectrum analysis
code {\sc detail/surface} \citep{Giddings81, ButlerGiddings85}. Later
on, \citet{MartinsHillier12} explored the formation of \CIII\,4647-50-51 
and \CIII\,5696 in detail, and found a tight coupling of
these lines to UV-transitions that regulate the population of the
associated levels. 

Though data and observations improved with time, some ``classical''
problems are still discussed and partly an issue even to-date,
particularly regarding the establishment of a consistent ionization
equilibrium for \cions: Often, \CII\ provides solar, but also
sub-solar abundances in early-type main sequence stars
(\citealt{daflon99} or \citealt{daflon01a}); \CIII\ might display
solar abundances in OB dwarf stars and O supergiants
(\citealt{daflon01b} and \citealt{pauldrach01}); and \CIV\ can lead to
all sorts of results, mostly because of the very restricted number of
lines (four in the optical, but only in the rarely observed range
between 5,000 and 6,000\,\AA; furthermore, two of these four lines are
weak, and seldom, if at all, discussed/analyzed).
Differences in abundance from \CII\ vs. \CIII\ can reach a factor of 5 to
10 \citep{hunter07}, and even when considering \CII\ lines
alone, there can be significant line-by-line variations. 
In B4-O6 stars,
\CII\,4267 might indicate a very low abundance when compared with
weaker lines such as the doublet at 6578-6582\,\AA\ (\citealt{kane80}).

Recent studies have called the attention to the importance of 
implementing precise atomic data when some of these classical problems
are addressed: Problematic data might produce systematic
discrepancies in the final results, independent of the specific
atmospheric model used, since these data describe interactions
governed by the laws of quantum mechanics, independent of their
environment.

On the other hand, this dependence can be used to test specific atomic
models regarding their capability to reproduce the observed spectral
features. Prior to this final proof of reliability, though, a series of
tests should be performed, including a comparison with alternative
models, in order to investigate the impact of the various components
of the model atom on the final result. For our purpose, the
atmospheres of late to early O-type stars represent suitable testbeds,
because within this temperature range the main ionization stage of
carbon changes drastically. Therefore, a grid of representative O-type
stars permits us to examine the quality of the results produced by our
newly developed carbon model atom.

Obviously, a spectroscopic analysis does not only depend on the atomic
data and the atmospheric model, but also on the quality of the
observational data. This even more for the tests outlined above: High
S/N spectra are needed, preferably from slowly rotating single
stars. The projected rotational velocity, \vsini, is one of the major
broadening agents capable of making the majority of carbon lines
almost invisible in the entire optical spectrum, recognizing that
these are mostly weak lines (\citealt{wolff82}). \vsini\ also affects
the blending of a set of diagnostic lines by lines from other atoms, 
(e.g., the strong \CIII\,4647-50-51 complex blended by many \OII\ lines).

As already pointed out, the optical diagnostics of carbon in O-type
stars is also influenced by a variety of UV transitions. Thus, a
proper treatment of the UV radiation is necessary, both for the
optical analysis and for an independent or combined investigation of
UV carbon lines. If at least part of these lines are formed in the
wind, the inclusion of X-rays and extreme ultraviolet (EUV) emission
from wind-embedded shocks becomes essential. As a first step of this
complex analysis, we can identify those optical lines that have levels
pumped by UV transitions, and investigate how strong the radiation
from wind-embedded shocks must be to influence the line shapes
significantly.

Besides the X-ray emission, the UV region is also influenced by 
micro- and macro-clumping, and porosity in velocity space, which makes
the analysis even more complex. This issue, however, will be addressed
in a forthcoming study, after we have convinced ourselves of the
reliability of our carbon model.

This paper is organized as follows. Sect.~\ref{section_2} summarizes important
characteristics of our atmosphere code, {\sc FASTWIND}, and details
our newly developed carbon model atom, the set of diagnostic lines
used, and the model grid adopted as testbed. In Sect.~\ref{tests} we provide
various tests performed to check our model atom. Sect.~\ref{results} presents all
relevant results from comparing synthetic carbon spectra with observed
ones, for the case of six slowly-rotating O-type stars of various
spectral type and luminosity class. Moreover, we discuss the potential
impact of X-ray/EUV radiation from wind-embedded shocks on the
optical carbon lines. In Sect.~\ref{summary} we conclude with an overview of the
present work as the basis for a more detailed future analysis. 

\section{Prerequisites for a carbon diagnostics} 
\label{section_2} 

All the calculations described in this work have been performed with
the latest update (v10.4.5) of the NLTE model atmosphere/spectrum
synthesis code {\sc FASTWIND} \citep{puls05, rivero12}. It includes
the recent implementation of emission from wind-embedded shocks and 
related physics, which will be used here to investigate potential effects of
X-rays/EUV radiation on the selected optical carbon lines. A detailed
description of the X-ray implementation in {\sc FASTWIND} is given by 
\citet{carneiro16}.

\subsection{The code} 
\label{the_code} 

For the diagnostics of early-B and O-type stars, {\sc FASTWIND} thus
far used models atoms for H, He, N (developed by \citealt{puls05} and
\citealt{rivero122}), Si (see \citealt{Trundle04}), while data for C,
O, and P have been taken from the WM-{\sc basic} database
\citep{pauldrach01}. We call these elements `explicit' (or
foreground) elements. Briefly\footnote{For a more detailed
description of {\sc FASTWIND}, and the philosophy of explicit and
background elements, see \citet{puls05} and \citet{rivero11}.}, such
foreground elements are used as diagnostic tools and treated with high
precision by detailed atomic models and by means of comoving frame radiation 
transport for all line transitions. Most of the other elements up to
Zn are treated as so-called background elements. Since these are
necessary `only' for the line-blocking/blanketing calculations, they
are treated in a more approximate way, using parameterized ionization
cross-sections in the spirit of \citet{seaton58}. Only for the most important
lines from background elements, a comoving frame transfer is performed,
while the multitude of weaker lines is calculated by means of the
Sobolev approximation. The latter approximation is applicable for the wind
regime, but it may fail for regions with a curved velocity field
(transition between photosphere and wind!), and in the deeper
photosphere. The Sobolev approximation, when applied to regions with a
pronounced velocity field curvature, yields too highly populated upper
levels in line transitions (see, e.g, \citealt{santo97}). This could
directly affect our carbon analysis, and is one more reason to use
carbon as an explicit element and to develop a corresponding, more
detailed carbon model.  



\subsection{The carbon model atom}
\label{carbon_model}

The first step regarding the development of a new model atom 
concerns the decision of how many and which states shall be included 
into each ion. We established a sequence of criteria to define our 
choice of levels. At first, as suggested by \cite{hubeny98}, the gap 
of energy between the highest ion level and the ground state of the next 
ionization stage should be less than $kT$. Since our conventional
O-star grids include a minimum \Teff\ of $\sim$28~kK, this temperature 
was chosen to establish a first guess for the uppermost levels of
\CIII\ and \CIV. In the case of \CII, we used a temperature of
22~kK to obtain a better representation of this ion in B stars.

With a first list of levels, the second criterium was to account for
{\it all} levels within a given subshell, up to and including the subshell
considered by criterium one, which extends our previous list by a few
more levels. Subsequently, a third and final criterium was to re-check
the Grotrian diagram and to include higher lying levels with multiple
transitions downward.  

At this point, the uppermost considered level has an energy far beyond
the limit established by the first criterium. Even though, the second
criterium was revisited for completeness, and few more levels (partly
with very weak cross-sections) included as a final step. 

Basically, the list/configuration and energies of levels were taken
from NIST\footnote{http://www.nist.gov/physlab/data/asd.cfm, described
in \citet{nist}}, but we cross-checked with other databases relying on
independent calculations. In particular, the list of levels used in
this work agrees to a large part with the WM-{\sc basic}
database~\footnote{{See \citet{pauldrach94c}. Briefly, the atomic
structure code {\sc superstructure} (\citealt{eissner69, eissner91})
has been used to calculate all bound state energies in LS and
intermediate coupling as well as related atomic data, particularly
oscillator strengths including those for stabilizing transitions.}}
and also with the OPACITY Project online database~
\footnote{http://cdsweb.u-strasbg.fr/topbase/topbase.html} (TOPbase
hereafter, see \citealt{cunto92} for details). The order of levels may
appear, in few cases, interchanged in different databases, due to
slightly different energies. 

Oscillator strengths were mainly taken from NIST, though this
database only provides data for allowed transitions.
For a given radiative bound-bound transition, the {\it gf}-values are
very similar in the different databases inspected by us: NIST, WM-{\sc
basic}, and data from an application of the Breit-Pauli method
\citep{nahar02}. Data for forbidden transitions were essentially taken
from the WM-{\sc basic} database. Radiative intercombinations have
been neglected, because of negligible oscillator strengths. 

TOPbase displays photoionization cross-section data from calculations
by \citet{seaton87} for almost all the levels included in our model
atom. Already \citet{nieva08} presented a comparison between the
radiative bound-free data from TOPbase and \citet{nahar97}, concluding
that the use of TOPbase reproduces more accurately the \CII\,4267, 
6151 and 6462 transitions, which are also of our
interest. On the other hand, within the OPACITY Project no data were
calculated for highly excited terms (e.g., C2\_37: $^{2}$G or C2\_38:
$^{2}$H$^{0}$, see Table \ref{levels_cii}), because the quantum defect
is zero, which means that such levels can be approximated as hydrogen-like. 
For these cases, we used the resonance-free cross-sections
provided in terms of the \citet{seaton58} approximation 
\beq
\alpha(\nu) = \alpha_{0}[\beta(\nu_{0}/\nu)^{s} + (1-\beta)(\nu_{0}/\nu)^{s+1}], 
\label{seaton_aprox} 
\eeq 
with $\alpha_{0}$ the threshold cross-section at $\nu_0$, and $\beta$ and
$s$ fit parameters, all taken from the WM-{\sc basic} database. 

The radiative bound-free data from TOPbase, which is our primary
source, include the numerous complex resonance transitions relevant
for the description of dielectronic recombination and reverse
ionization processes. For the few levels where no data are present
(see above), we used the ``explicit'' method accounting for individual
stabilizing transitions (see, e.g., \citealt{rivero11}), with data
from WM-{\sc basic} (a further discussion on this approach
will be provided in Sect.~\ref{tests}).

Collisional ionization rates are calculated following the
approximation by \citet{seaton62}. The corresponding threshold 
cross-sections are taken from WM-{\sc basic} and \citet{nahar02},
which present similar values for the majority of levels, and these
also in agreement with TOPbase.

For collisional excitations, we used a variety of suitable data-sets, 
discussed in the following together with particularities for each carbon ion:

{\bf \CII} is described by 41 LS-coupled levels, roughly up to principal 
quantum number $n = 7$ and angular momentum $l= 5$, with all
fine-structure levels being packed\footnote {To calculate the final
synthetic profiles by means of the formal integral, these levels are
unpacked. To this end, we assume that $n_{\rm i}/g_{\rm i}$
(occupation number divided by statistical weight) is similar
within each of the sublevels belonging to a packed level, due to
collisional coupling.}. These levels are displayed in
Table~\ref{levels_cii}. For the 16 lowermost levels of this boron-like
ion, effective collision strengths were taken from R-matrix
computations by \citet{wilson05,wilson07}. For the remaining
transitions without detailed data, collisional excitation is
calculated using the \citet{vanregemorter62} approximation for optically
allowed transitions, and by means of the \citet{allen73} expression for the
optically forbidden ones. For the latter, corresponding collision strengths
$\Omega$ vary from 0.01 ($\Delta n \geq 4$) to 100 ($\Delta n =0$).
Over 300 radiative and 1000 collisional transitions have been included. 

{\bf \CIII} consists of 70 LS-coupled levels, until $n = 9$ and $l = 2$, with
fine-structure levels being packed. The levels are detailed in 
Table~\ref{levels_ciii}.  For electron impact excitation of the lowest
24 levels, we used the Maxwellian-averaged collision strengths
calculated by \citet{mitnik03} through R-matrix computations. The
collisional bound-bound data for the other levels were treated in
analogy to corresponding \CII\ transitions. This Be-like ion
comprises approximately 700 radiative and 2000 collisional transitions.

{\bf \CIV} includes 50 LS-coupled terms, until $n = 14$ and $l = 2$, with
fine-structure levels again being packed, and described in
Table~\ref{levels_civ}. \citet{aggarwal04} provide electron impact
excitation data for the lowest 24 fine-structure levels, which have been
added up in such a way as to be applicable for our first 14 terms. All remaining
collisional bound-bound transitions were treated in analogy to \CII.
Overall, this Li-like ion is described by roughly 200 radiative and 1000
collisional transitions. 

Thus far, {\bf \CV} consists of only one level, the ground state
(C5\_1: 1s$^{2}$ $^{1}$S), required for ionization/recombination
processes from and to \CIV. Anyhow, this is a suitable description, 
since (i) a further ionization is almost impossible under
O-star conditions, due to a very high ionization energy, and (ii) the
excitation energies of already the next higher levels are also quite
large, so that \CV\ should remain in its ground state.

\begin{table*}
\center 
\caption{Diagnostic carbon lines in the optical spectra of
early B- and O-type stars, together with potential blends. Lines with
wavelengths written in bold-face are visible in a wide temperature
range, and quite sensible to carbon abundance variations. Therefore,
they are most important for an optical carbon analysis (see also
Sect.~\ref{set_lines}). The
labels displayed in column 3 (``transition'') are detailed in
Tables~\ref{levels_cii}, \ref{levels_ciii}, and \ref{levels_civ}.}
\label{tab_lines_carbon}
\begin{tabular}{ccccccc}
\hline 
\hline
Ion & Wavelength(\AA) &Transition & $f$ & log($gf$) & Blends & Comment \\
\hline
\CII & {\bf 3918.98} &C2\_7 - C2\_11  & 0.1470 & -0.533 & \NII\  $\lambda$3919.00, \OII\ $\lambda$3919.2  & weak line \\ 
     & {\bf 3920.69} &C2\_7 - C2\_11  & 0.1460 & -0.232 & \SIII\ $\lambda$3920.29& " \\
     & {\bf 4267.00} &C2\_9 - C2\_16  & 0.9140 &  0.563 & \SII\  $\lambda$4267.76& strong \CII\ doublet \\
     & {\bf 4267.26} &C2\_9 - C2\_16  & 0.8670 &  0.716 & \FeII\ $\lambda$4267.82& " \\
     & {\bf 4267.26} &C2\_9 - C2\_16  & 0.0434 & -0.584 & \BaII\ $\lambda$4267.82& " \\
     & 4637.63&C2\_12 - C2\_27 & 0.0295 & -1.229 & \FeI\  $\lambda$4637.50& region
dominated by \OII $\lambda$4638.9 \\
     & 4638.91&C2\_12 - C2\_27 & 0.0266 & -0.973 & \SiIII\ $\lambda$4638.28 & " \\ 
     & 4639.07&C2\_12 - C2\_27 & 0.00295& -1.928 & \TiI\  $\lambda$4639.94 & " \\
     & {\bf 5132.94} &C2\_13 - C2\_33 & 0.3070 & -0.211 & - & weak doublet \\
     & {\bf 5133.28} &C2\_13 - C2\_33 & 0.1660 & -0.178 & - & " \\
     & 5139.17&C2\_13 - C2\_33 & 0.0491 & -0.707 & -             & visible in slow rotators \\
     & {\bf 5143.49} &C2\_13 - C2\_33 & 0.1530 & -0.212 & - & " \\
     & {\bf 5145.16} &C2\_13 - C2\_33 & 0.2580 &  0.189 & - & " \\
     & 5151.09&C2\_13 - C2\_33 & 0.1100 & -0.179 & - & " \\
     & 5648.07&C2\_13 - C2\_31 & 0.0943 & -0.249 & \FeII\ $\lambda$5648.89& not visible \\
     & 5662.47&C2\_13 - C2\_31 & 0.0939 & -0.249 & \TiI\ $\lambda$5662.14, \FeI\ $\lambda$5662.51 & weak line \\
     & 6151.53&C2\_14 - C2\_28 & 0.0049 & -1.310 & - & not visible \\
     & 6461.94&C2\_16 - C2\_29 & 0.1150 &  0.161 & \MnII\ $\lambda$6462.21 & " \\
     & {\bf 6578.05} &C2\_6  - C2\_7  & 0.7140 &  0.154 & - & weak doublet \\
     & {\bf 6582.88} &C2\_6  - C2\_7  & 0.2380 & -0.323 & \NII\ $\lambda$6582.60 & " \\
\hline
\CIII& {\bf 4056.06}&C3\_24 - C3\_44 & 0.3700 &  0.267 & \TiI\ $\lambda$4055.01, \MnI\ $\lambda$4055.54  & strong \\ 
     & 4068.90&C3\_20 - C3\_39 & 0.9830 &  0.838 & \ScIII\ $\lambda$4068.66, \OII\ $\lambda$4069.62& " \\ 
     & 4070.26&C3\_20 - C3\_39 & 0.9960 &  0.953 & \OII\ $\lambda$4069.88                          & " \\
     & {\bf 4152.51}&C3\_23 - C3\_43 & 0.2580 & -0.112 & \NIII\ $\lambda$4152.13, \NeIII\ $\lambda$4152.58 & " \\
     & {\bf 4156.50}&C3\_23 - C3\_43 & 0.2290 &  0.059 & \LiII\ $\lambda$4156.45,\FeI\ $\lambda$4156.67 & strong doublet  \\
     & {\bf 4162.86}&C3\_23 - C3\_43 & 0.2360 &  0.218 & \SVI\ $\lambda$4162.28,\SII\ $\lambda$4162.66 & " \\
     & {\bf 4186.90}&C3\_22 - C3\_40 & 1.1800 &  0.918 & \FeI\ $\lambda$4187.03, \FeI\ $\lambda$4187.59 & X-ray dependent \\
     & {\bf 4647.42}&C3\_7 - C3\_10  & 0.3920 &  0.070 & \SII\ $\lambda$4648.17  & UV-transition dependent \\
     & {\bf 4650.25}&C3\_7 - C3\_10  & 0.2350 & -0.151 & \TiI\ $\lambda$4650.01, \OII\ $\lambda$4650.84  & " \\
     & {\bf 4651.47}&C3\_7 - C3\_10  & 0.0783 & -0.629 & " &" \\
     & {\bf 4663.64}&C3\_13 - C3\_26 & 0.0984 & -0.530 & \AlII\ $\lambda$4663.05 & strong doublet \\
     & {\bf 4665.86}&C3\_13 - C3\_26 & 0.2210 &  0.044 & \SiIII\ $\lambda$4665.87 & " \\
     & {\bf 5253.57}&C3\_13 - C3\_25 & 0.0654 & -0.707 & \FeII\ $\lambda$5253.46 & weak line \\
     & {\bf 5272.52}&C3\_13 - C3\_25 & 0.0653 & -0.486 & \NV\ $\lambda$5272.18, \FeII\ $\lambda$5372.22 & X-ray dependent \\
     & {\bf 5695.92}&C3\_9 - C3\_12  & 0.3460 &  0.017 & \AlIII\ $\lambda$5696.60 & UV-transition dependent \\
     & 5826.42&C3\_24 - C3\_34 & 0.5220 &  0.417 & \FeII\ $\lambda$5826.52 & weak line \\
     & 6731.04&C3\_13 - C3\_23 & 0.1700 & -0.293 & \OIII\ $\lambda$6731.13 & not visible \\
     & 6744.38&C3\_13 - C3\_23 & 0.1900 & -0.022 & -                     & X-ray dependent \\
     & {\bf 8500.32}&C3\_8 - C3\_9   & 0.3280 & -0.484 & \FeII\ $\lambda$8499.61 & " \\
\hline
\CIV & {\bf 5801.33}&C4\_3 - C4\_4   & 0.3200 & -0.194 & - & X-ray dependent \\ 
     & {\bf 5811.98}&C4\_3 - C4\_4   & 0.1600 & -0.495 & \OII\ $\lambda$5011.79 & " \\ 
     & 5016.62&C4\_11 - C4\_15 & 0.1750 & -0.456 & \ArII\ $\lambda$5016.47 & weak line \\ 
     & 5018.40&C4\_11 - C4\_15 & 0.1750 & -0.155 & \FeII\ $\lambda$5018.43 & " \\ 
\hline 
 \end{tabular}
\end{table*}

To summarize, our carbon model atom comprises 162 LS-coupled levels,
basically ordered following NIST. In few cases, we interchanged the
order and adapted the corresponding energies, to obtain a compromise
with the level-lists from WM-{\sc basic} and TOPbase, which have been
used for a large part of bound-bound and the majority of bound-free
data, respectively. We note that such a task has to be done with
specific care, since any wrong labeling would lead to spurious results.
The definition of \cionss\ accounts all together for more than 1000
radiative and 4000 collisional transitions. 


\subsection{Diagnostic optical carbon lines}
\label{carbon_lines}

We selected a set of 43 carbon lines visible (at least in principle)
in the optical spectra of OB-stars, which allow us to approach some of
the ``classical'' problems already mentioned in Sect.~\ref{Introduction}, as for
example: (i) inconsistent carbon abundances implied by \CII\,4267 
and \CII\,6578-82 (\citealt{grigsby04},
\citealt{hunter07}), (ii) abundances derived from \CII\ and \CIII\ may
differ by a factor of ~5-10 (\citealt{daflon01b}, \citealt{hunter07}),
(iii) the difficulty to establish a consistent ionization equilibrium
for \cions\ (\citealt{nieva06,nieva07,nieva08}). 

NIST identifies all relevant lines in the spectrum, together with 
corresponding oscillator strengths. This was our first source for building 
a prime sample of lines. We inspected various observed spectra 
(partly described below) to identify which of these lines are blended, and to
find additional lines not included so far. In the end, we defined a
set of lines similar to the ones used by \citet{nieva08}, with some
relevant additions. For the final synthetic spectra, we adopt
Voigt profiles, with central wavelengths from NIST, 
radiative damping parameters from the Kurucz database\footnote{e.g.,
www.pmp.uni-hannover.de/cgi-bin/ssi/test/kurucz/sekur.html}, and
collisional damping parameters computed according to \cite{Cowley71}.

Table~\ref{tab_lines_carbon} presents three different blocks, divided
into \CII, \CIII, and \CIV. The second column displays the wavelengths
of the lines, followed by the lower and upper level of the considered
transition. Columns 4, 5 and 6 display the oscillator strengths, the
$\log(gf)$-values, and potential blends. The last column provides a
short comment relevant for each line.

\subsection{Model grid}
\label{model_grid}

In this study, we use the ``theoretical'' O-star model grid originally
designed by \citet[their Table 5]{pauldrach01}\footnote{This grid, in
turn, is based on observational results from \citet{Puls96}, which at
that time did not include the effects of wind inhomogeneities, so that
the adopted mass-loss rates might be too large, by factors from
$\sim$3~{\ldots}~6.}, revisited by \citet{puls05} to compare results
from an earlier version of {\sc FASTWIND} with the outcome of
WM-{\sc basic} calculations, and again revisited by \citet{carneiro16}
to test our recently developed X-ray implementation.
Table~\ref{tab_grid} displays the stellar and wind parameters of the
grid models. The adopted models allow us to study, for a certain
range of spectral types, how changes in stellar parameters (e.g,
\Teff, \logg, carbon abundance) will affect the shape and strength of
significant carbon lines. At the same time, these models define a reasonable
testbed for a series of tests described in Sect.~\ref{tests}.

We adopt solar abundances from \cite{asplund09}, together with a
helium abundance, by number, $N_{\rm He}$/$N_{\rm H}$ = 0.1. 
Carbon abundances different from the solar value are explicitly
mentioned when necessary. 

The main focus of this work is set on the analysis of photospheric
carbon lines, which should not be affected by wind clumping. In
the scope of this work, we thus only consider homogeneous wind models.
Even though, our unclumped
models with mass-loss rate \mdot$_{uc}$ {\it roughly} correspond to
(micro-)clumped models with a lower mass-loss rate, \mdot$_{c}$, 
\beq 
\mdot_{c} =\mdot_{uc}/\sqrt{\fcl},
\label{mdotclump} 
\eeq 
where $\fcl \ge 1$ is the considered clumping factor.  


\begin{table}
\tabcolsep1.7mm
\begin{center}
\caption{Stellar and wind parameters of our grid models with
homogeneous winds, following \citet{pauldrach01}. For all models, the
velocity field exponent has been set to $\beta$= 0.9, and a micro-turbulent
velocity, $\vturb$ = 15~\kms, has been used.
}
\label{tab_grid}
\begin{tabular}{lllcrc}
\hline
\hline
\multicolumn{1}{l}{Model}
&\multicolumn{1}{l}{\Teff}
&\multicolumn{1}{l}{\logg}
&\multicolumn{1}{c}{\Rstar}
&\multicolumn{1}{c}{\vinf}
&\multicolumn{1}{c}{\mdot}
\\
\multicolumn{1}{l}{}
&\multicolumn{1}{l}{\tiny{(kK)}}
&\multicolumn{1}{l}{\tiny{(\cms)}}
&\multicolumn{1}{c}{\tiny{(\rsun)}}
&\multicolumn{1}{c}{\tiny{(\kms)}}
&\multicolumn{1}{c}{\tiny{(\mdu)}}
\\
\hline
\multicolumn{6}{c}{Dwarfs} \\
\hline
\vspace{0.07mm}
D30 &  30  & 3.85   &  12  &   1800   &  0.008  \\
D35 &  35  & 3.80   &  11  &   2100   &  0.05   \\
D40 &  40  & 3.75   &  10  &   2400   &  0.24   \\
D45 &  45  & 3.90   &  12  &   3000   &  1.3    \\
D50 &  50  & 4.00   &  12  &   3200   &  5.6    \\
D55 &  55  & 4.10   &  15  &   3300   &  20     \\
\hline
\multicolumn{6}{c}{Supergiants} \\
\hline
\vspace{0.07mm}
S30 &  30  & 3.00   &  27  &   1500   & 5.0   \\
S35 &  35  & 3.30   &  21  &   1900   & 8.0   \\
S40 &  40  & 3.60   &  19  &   2200   & 10    \\
S45 &  45  & 3.80   &  20  &   2500   & 15    \\
S50 &  50  & 3.90   &  20  &   3200   & 24    \\
\hline
\end{tabular}
\end{center}
\end{table}

\subsection{Observational data}
\label{obs_data}

In Sect.~\ref{results}, we will use optical spectra (kindly provided by Holgado et
al. 2017, submitted to A\&A) from prototypical O-type stars, to
compare with the carbon line profiles as calculated using our new
model atom. These stars are included in the grid of O-type standards,
as defined in \citet{Maiz15}\footnote{covering 131 Galactic stars in the
spectral range from O2 to O9.7 (all luminosity classes) in the
Northern and Southern hemisphere.}. From the observed sample, we
selected six presumably single stars in different ranges of
temperature and with low \vsini. The spectra have
been collected by means of three different instruments: HERMES
(with a typical resolving power of $R$ = 85,000, see \citealt{raskin04})
at the MERCATOR 1.2\,m
telescope, FEROS ($R$ = 46,000, see \citealt{kaufer97}) at the
ESO 2.2\,m telescope,
and FIES ($R$ = 46,000, see \citealt{telting14}) at the NOT 2.6\,m 
telescope.
Table~\ref{tab_spectra} lists the instrument and S/N ratio of each
spectrum analyzed in this work. More details are provided in
Sect.~\ref{stellar_sample}.


For the temperature range considered in this work, we expect that
carbon line profiles from ionizations stages \cions\ are visible around
$\sim$30~kK. On the other hand, for the hottest objects ($\sim$50~kK),
we will have to rely on estimates using \CIV\ lines alone.

\section{Testing the atomic model}
\label{tests}

After having constructed a new carbon model atom using high quality
data, this section describes some of the tests we performed to
investigate the outcomes from using this model atom in an atmospheric
code, for various stellar conditions. Specific tests are briefly
summarized in the following:

(i) As outlined in Sect.~\ref{the_code}, previous {\sc FASTWIND}
calculations used the carbon model atom from the WM-{\sc basic}
database, independent of whether carbon was treated as a foreground or
background element. Thus we are able to compare the results from our
former practice and our new (and more detailed) description. As
expected, in terms of ionization fraction, both methods display
exactly the same results in the stellar photosphere. Irrespective of
wind-strength, differences appear only in the outer wind (e.g., for
model S30, around $\taur \leqslant 10^{-4}$, corresponding to $r
\geqslant$ 6~\Rstar\ or $v(r) \geqslant$ 0.8~\vinf) for all considered
ions except for \CII, for which differences begin to appear deeper in
the wind (again for model S30, around $\taur \leqslant 10^{-2}$,
corresponding to $r \geqslant$ 3~\Rstar\ or $v(r) \geqslant$
0.2~\vinf). Our new carbon description displays consistently less
\CII\ for a wide range of temperatures (for both dwarf- and
supergiant-models), where the maximum difference
(0.7~dex) is reached in our coolest model at 30~kK. This behaviour is 
due to less \CIII\ and \CIV (see below), though the differences for these ions are
lower (less than 0.5~dex), and appear only in the outer wind. 

(ii) In our model atom, we use the expression from \citet{allen73}, 
with individual $\Omega$ values from 0.01 to 100, to describe those
collisional bound-bound transitions where the radiative ones are
forbidden and where we lack more detailed data (usually, between quite
highly excited levels). We tested the impact of uncertainties in
$\Omega$ on the final results, by setting $\Omega =1.0$ for all these
transitions, and found that this has a negligible impact on our
results regarding the optical lines. Indeed, the ``exact'' value of the
collisional strength is only important for a specific part of the
atmosphere in between the LTE regime and the much lesser dense wind.
Since we use Allen's expression only for those transitions where the
radiative ones are forbidden, i.e., which have a very low oscillator
strength ($\leqslant 10^{-5}$), the weak impact of $\Omega$ is
understandable when considering the dominating effect of the other
radiative transitions included in the model atom.  We expect, however,
that specific IR-transitions might be influenced though. 

(iii) We also tested a possible interplay between nitrogen and carbon,
which might arise when combining different foreground elements in 
{\sc FASTWIND}. To this end, we considered three different model series: 
one with H/He + carbon + nitrogen as foreground elements, one with
H/He + only carbon, and one with H/He + only nitrogen. In the latter
two cases, either nitrogen or carbon are used as background elements,
respectively, with atomic data from WM-{\sc basic}. These tests
resulted in irrelevant differences regarding the carbon ionization
stratification ($\sim$~0.1~dex in the outer wind), when comparing the
HHeCN and the HHeC models. The same, now regarding nitrogen, holds
when comparing HHeCN vs. HHeN: we found no visible difference in the
nitrogen description, whether carbon is included or not. We emphasize
though that this test does not consider potential C/N line overlap
effects, particularly regarding the EUV resonance lines from C and N at 
$\sim$321\,\AA
\footnote{similar to the case of overlap between N and O resonance
lines at $\sim$374\,\AA,
which is decisive for the formation of \NIII\,4634-40-42 (see
\citealt{rivero11}).}. This issue deserves a separate investigation. 

These first tests confirmed our expectations, illuminating specific 
aspects that have low interference on the final results. Of course, 
we have tested our model atom much more. In the following, some of these 
tests are discussed in more detail.

\subsection{Dielectronic Recombination}
\label{tests_dr}

One advantage for testing our carbon description is the availability
of two independent codes in our scientific group ({\sc FASTWIND} and
WM-{\sc basic}), which can be used to calculate the same atmospheric
models but employing different atomic models. A comparison
of the carbon ionization stratification then, for a set of models
calculated with {\sc FASTWIND} and WM-{\sc basic}, gives a quick
overview about differences between our results and former work (see
\citealt{pauldrach94c, pauldrach01}). 

In this spirit, we calculated all grid models described in
Table~\ref{tab_grid} also with WM-{\sc basic}. After
comparing these models with corresponding {\sc FASTWIND} ones, we found
a rather similar run of \CIV\ and \CV, both in the stellar photosphere
and also in the wind. In contrast, \CII\ and \CIII\ displayed a
recurrent difference for all the models: in the wind part, our results lay
consistently one or two dex below the outcome from WM-{\sc basic}.
%
%
Though this finding does not allow for premature conclusions (at least
at this stage, we do not know what is the better description),
it nevertheless caught our attention, especially since the same
discrepancy had been found for a wide range of temperatures. We thus
recalculated the {\sc FASTWIND} models, but this time using the
complete WM-{\sc basic} dataset for carbon. Comparing with our initial
models, we found the same difference in \CII\ and \CIII\ as described
just above. Thus the differences need to be attributed to the
different datasets and not to the different atmospheric models, and we
set out to compare both datasets in detail.

\begin{figure}[t]
\resizebox{\hsize}{!}
{\includegraphics[angle=90]{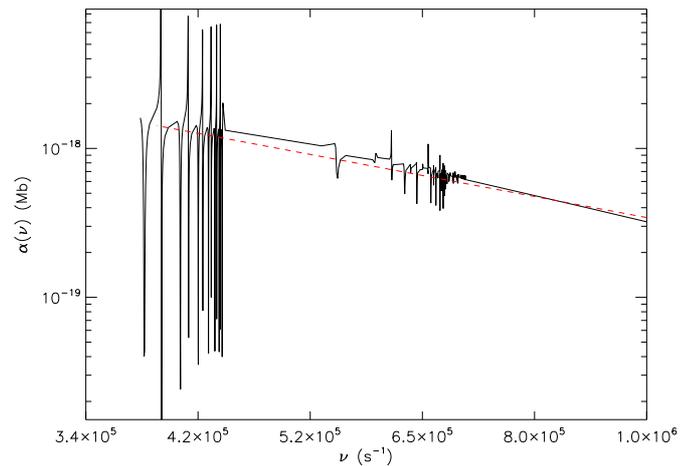}}
%
\caption{Bound-free cross-section of the \CIII\ ground state including
resonances (from OPACITY Project, used in our new model atom, black
line), and the resonance-free data (from WM-{\sc basic}, red line).} 
\label{crossbf}         
\end{figure}

In the end, we identified the origin of the discrepancy within the radiative
bound-free transitions, where each of both datasets describes these 
transitions differently. While within our new model atom we use an
{\it implicit method} to define the dielectronic recombination
(henceforth DR\footnote {Dielectronic recombination can be summarized
as ``the capture of an electron by the target leading to an
intermediate doubly excited state that stabilizes by emitting a photon
rather than an electron'' \citep{rivero11}.}) data within the
photoionization cross-sections, the WM-{\sc basic} database adopts an
{\it explicit method}. Both methods are implemented into {\sc
FASTWIND}: Within the {\it implicit method}, the resonances appear
``naturally'' in the photoionization cross-sections (from OPACITY
Project data, \citealt{cunto92}), whereas the {\it explicit method}
considers explicitly the stabilizing transitions from autoionizing
levels together with the resonance-free cross-sections. As an example,
Fig.~\ref{crossbf} displays the data available from the OPACITY
Project (black line) with the numerous complex resonances for the
ground state of \CIII, together with the \citet{seaton58}
approximation using data from WM-{\sc basic} (red line), to which the
stabilizing transitions (data input: frequencies and oscillator
strengths) would need to be added.

For further details, and advantages and disadvantages of both methods, 
we refer to \citet{hilliermiller98} and \citet{rivero11}. The
important point with respect to this work is the following: Since in
the explicit method one defines each stabilizing transition by
corresponding data, we have the possibility to remove any of those
transitions by setting the corresponding oscillator strengths to a very
low value.

\begin{figure}[t]
\resizebox{\hsize}{!}
{\includegraphics{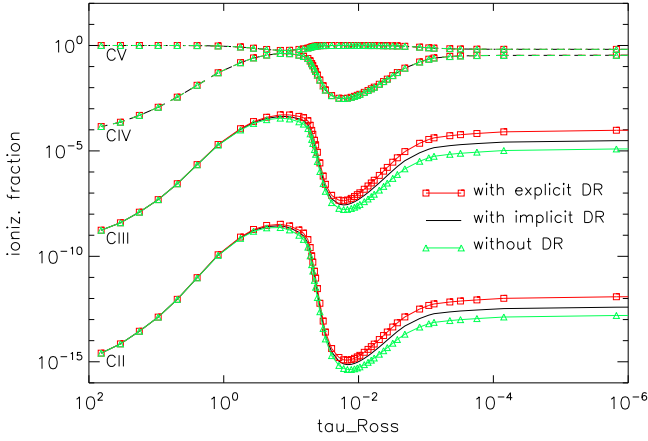}}
%
\caption{Ionization fractions of carbon ions, as a function of \taur,
for model D45, with and without dielectronic recombination (DR). Note
the impact of DR  onto \CII\ and \CIII\ in the wind region. For further
details, see text.}
\label{d45_dr}                                                  
\end{figure}

Fig.~\ref{d45_dr} shows the ionization fraction of different carbon
ions in the atmosphere, for model D45 (see
Table~\ref{tab_grid}). We calculated three different models, {\it where
only the bound-free dataset for carbon was changed}, leaving all other data at
their original value, defined by our new carbon model atom.
In the first model, we used the implicit method with bound-free
cross-sections from OPACITY Project data (black line), in the second we
used corresponding WM-{\sc basic} data (explicit method - red line),
and in the third model we discarded the DR-processes in the WM-{\sc
basic} data, i.e., used only the resonance-free cross-sections by
excluding all stabilizing transitions (green line). As displayed in
Fig.~\ref{d45_dr}, the effect of DR is irrelevant in the stellar
photosphere, where due to the high temperatures and densities the
``normal'' ionization/recombination processes dominate. In the wind
part, the impact of DR remains irrelevant for \CIV, but becomes
crucial for a precise description of \CIII. In the case of \CII, the
difference is mostly a consequence of changes in \CIII: 
Without DR, less ions are recombining from \CIV\ to \CIII, and thus
also from \CIII\ to \CII, due to the lower population. Thus, the
differences seen in \CIII\ are reproduced in \CII, whether DR is present
or not. Since \CIV\ is the main ionization stage, the slight increase
in its ionization fraction (without DR) is almost invisible.

All models described in Table~\ref{tab_grid} produce the same effect 
for \CIII\ and \CII\ when DR data are removed. Here we have
concentrated on model D45, since for this model we already investigated 
the effect of DR on the ionization of {\it oxygen} in a previous study 
\citep{carneiro16}. 

Of course, we investigated which transitions (regarding their lower
levels -- \CIII) are responsible for such a change in the wind
ionization. It turned out that almost all of the first 40 states are
involved, but that levels C3\_19, C3\_29, and C3\_30 (for
configuration and term designation, see Table~\ref{levels_ciii}) are
responsible for already half of the total effect, where these levels
ionize to the second state of \CIV.

Finally, we note that also the models calculated with WM-{\sc basic}
show the same reaction when DR is excluded (with respect to all or
individual stabilizing transitions). We conclude that the two codes
independently show a lower degree of \CII\ and \CIII, once DR is
neglected. On the other hand, when actually accounting for DR, the
detected differences can be attributed to different strengths of the
stabilizing transitions/resonances, where according to our tests all
recombining states are relevant, though specific transitions (see
above) have a particularly strong impact. As a last test on this
issue, we explicitly compared the strengths for the latter
transitions (see also \citealt{rivero11}, Sect.~A3), finding a
discrepancy of roughly a factor of two (with WM-{\sc basic} data 
providing larger values).


\begin{figure}[t]
\center
\resizebox{\hsize}{!}
{\includegraphics{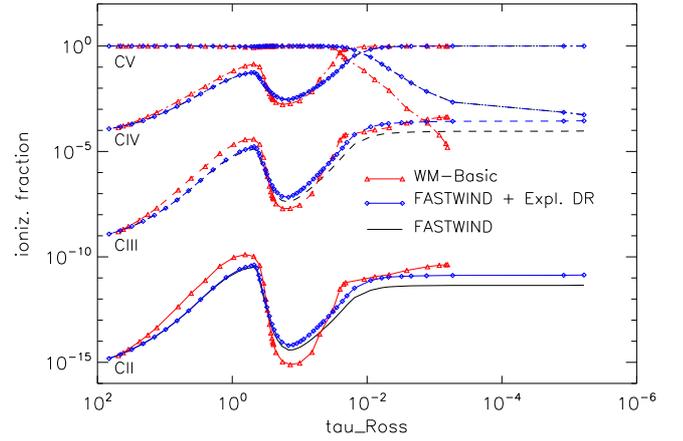}}

\caption{Ionization fraction of carbon ions, as a
function of \taur, for the S45 model, as calculated by WM-{\sc basic}
and {\sc FASTWIND} using different approaches for DR. For details, see
text.}
\label{wmbasic_s45}
\end{figure}

\subsection{Further comparison with WM-{\sc basic}}
\label{comp_models}

Once the importance of DR in transitions from \CIV\ to \CIII\ and its
indirect impact on \CII\ has been understood, we can continue
in our comparison between {\sc FASTWIND} and WM-{\sc basic} results.

We remind that both codes are completely independent (except that {\sc
FASTWIND} uses WM-{\sc basic} data for the background elements, i.e.,
for all elements different from H, He, and C in the case considered
here), and use different methods and assumptions. In addition to the
different treatment of metal-line blocking, WM-{\sc basic} calculates
the velocity field from a consistent hydrodynamic approach, leading to
certain differences particularly in the transonic region. Furthermore,
while WM-{\sc basic} uses the Sobolev approximation for all line
transitions and depths, {\sc FASTWIND} uses a comoving-frame transport
for the transitions from explicit elements and for the strongest lines
from the background ones. As already pointed out, this can lead to
significant discrepancies for those lines that are formed in the region
between the quasi-static photosphere and the onset of the wind.

Figure~\ref{wmbasic_s45} displays the comparison of ionization
fractions for carbon ions in the photosphere and wind (as a function
of \taur) for our S45 model (see Table~\ref{tab_grid}). Red lines
represent the carbon ionization stratification as derived by WM-{\sc
basic}, black lines show the {\sc FASTWIND} solution using our new
model atom, and blue lines display {\sc FASTWIND} models, where the
carbon bound-free transitions including DR are calculated using the
explicit method with WM-{\sc basic} data.  

For this grid model, \CIII\ and \CIV\ (the main ionization stage in the
wind) are of major relevance regarding a carbon line diagnostics,
though we also display \CII\ (irrelevant at this \Teff) and \CV,
approximated by only one ground-state level in our atomic model.
Within the photosphere, all solutions are quite similar, though at
certain depths differences become visible, mostly because of
deviations in the local velocity and density, and due to differences
in the line transfer (see above). In the wind, however, the standard
{\sc FASTWIND} and the WM-{\sc basic} solution diverge, not only for
S45, but also the other grid-models. This differences have been already
described in Sect.~\ref{tests_dr}, and are due to the different
description of DR. When we now manipulate our new model atom to use
the bound-free data from WM-{\sc basic} with their larger strengths
for the stabilizing transitions (blue curves), we indeed see much more
similar fractions also in the outer wind. 

In conclusion, we find a satisfactory agreement between results from
{\sc FASTWIND} and WM-{\sc basic}, if a similar treatment of DR is
performed. The differences apparent at first glance are due to the
fact that the stabilizing transitions in WM-{\sc basic} are larger
(or considerably larger for specific transitions) than implied by the
resonances provided by the OPACITY Project data.

Since we are no experts in this field, we cannot judge which data set
is the more realistic one, but until further evidence we prefer to use
the OPACITY Project data, since they are well documented, tested, and
applied within a variety of codes and studies.

\subsection{Optical carbon lines -- dependence on stellar parameters}
\label{para_dep} 

\begin{figure*}[t]
\center
{\includegraphics[scale=0.74,angle=90]{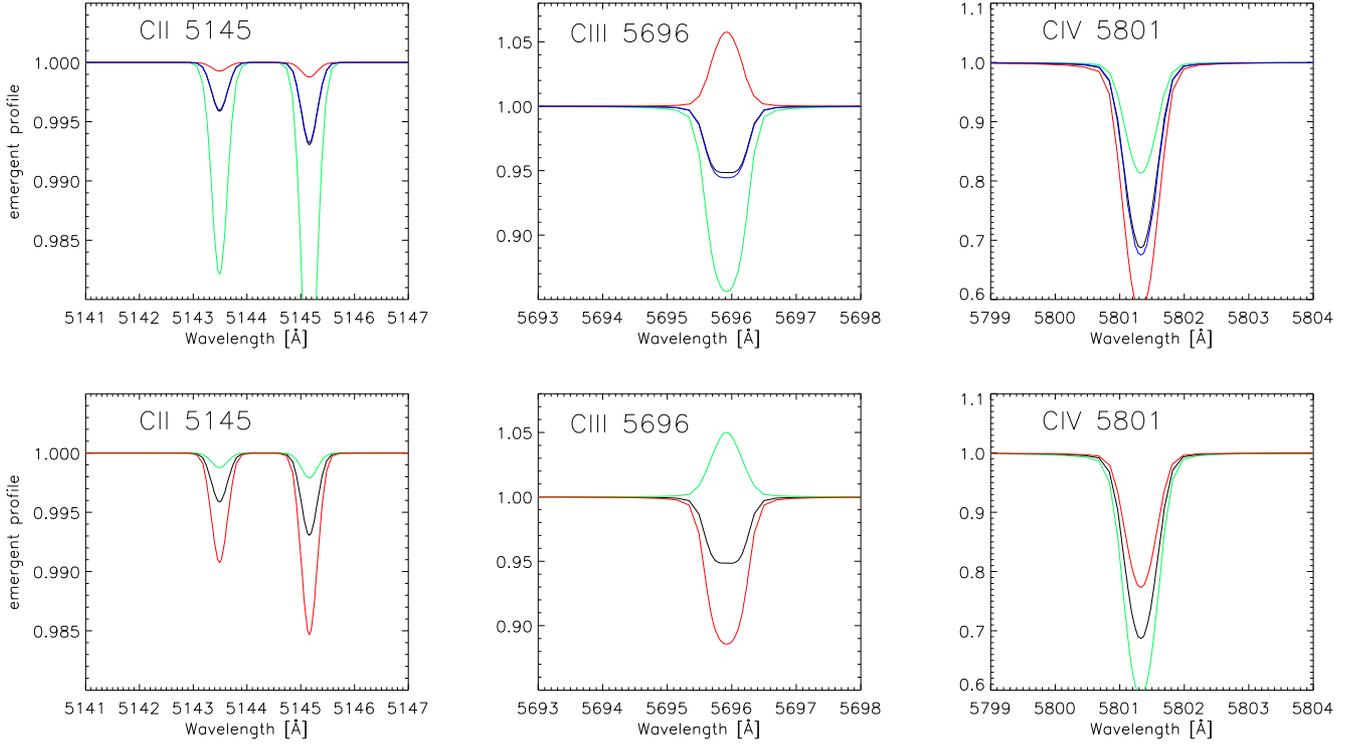}}
%
\caption{\CII\,5145, \CIII\,5696, and \CIV\,5801 line profiles for
model D35 (black lines) and similar models with relatively small
changes in effective temperature (\Teff) and gravity (\logg). In the
upper panels, the red lines correspond to a D35 model with \Teff\
increased by 1500~K, the green lines to a model with \Teff\ decreased
by the same value, while the blue lines display the reaction to a
decrease of \mdot\ by a factor of three. In the lower panels, the red
lines correspond to a D35 model with \logg\ increased by 0.2~dex, and
the green lines with \logg\ decreased by 0.2~dex.} 
\label{d35_cprof}
\end{figure*}

\vspace{0.5cm}

\begin{figure*}[h]
\center
{\includegraphics[scale=0.732,angle=90]{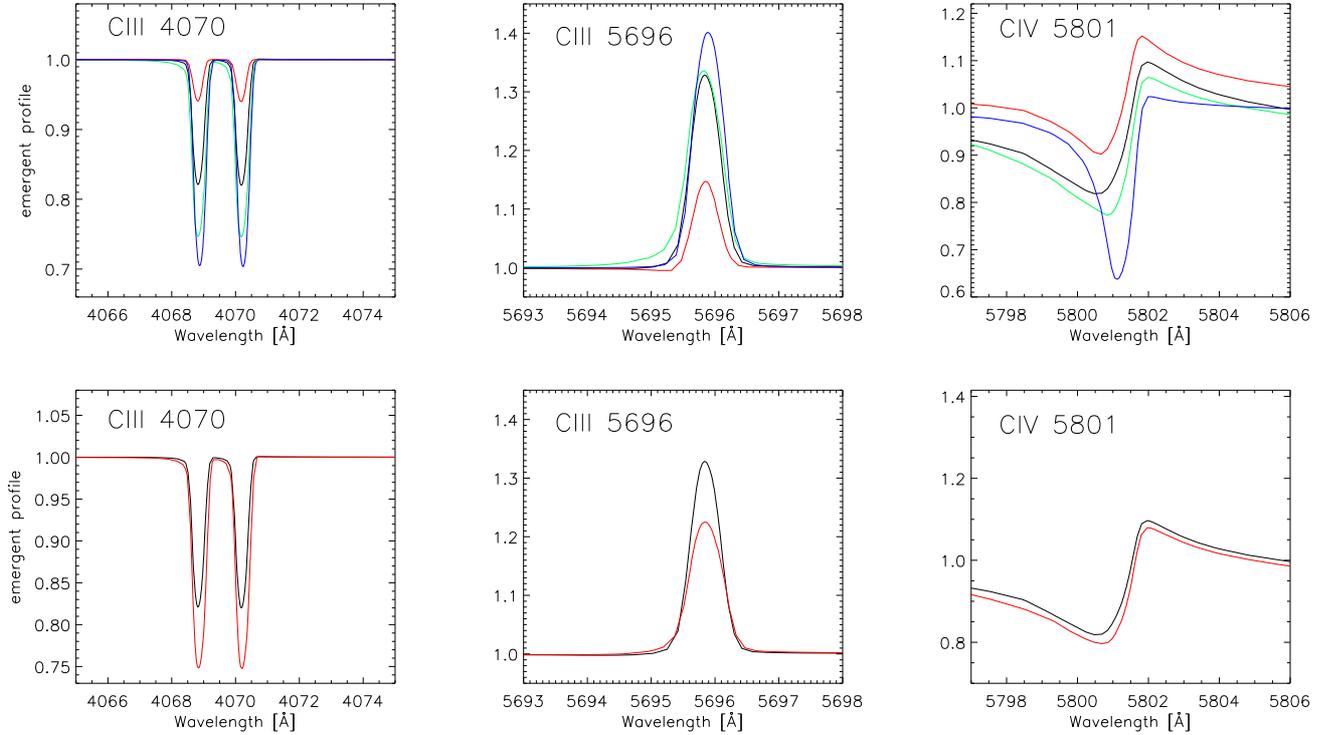}}
%
\caption{As Fig.~\ref{d35_cprof}, but for model S35. Since the \CII\
lines are absent in such a model, we display another strategic line
for this temperature range, \CIII\,4068-70. Note that \logg\ has only
been varied by +0.2~dex (red profiles) in the lower panel. See text
for details.} 
\label{s35_cprof}
\end{figure*}

The typical precision of a spectral analysis of massive stars using
H/He lines is on the order of $\pm$1.5~kK in effective temperature
and $\pm$0.1~dex in \logg\ (e.g., \citealt{repolust04}). Since we
aim at a non-LTE carbon abundance determination by line profile
fitting, we need to test the sensitivity of our set of strategic lines
to a variation of these parameters. 

Due to the distinct complexity in each line formation process, almost
each of the carbon lines will react differently.  Fig.~\ref{d35_cprof}
(and analogous figures) displays one spectral line per each carbon ion
in each of the three columns. The first column shows \CII\,5145, the
middle column \CIII\,5696, and the third one \CIV\,5801. These lines
have been chosen because they are strong (highlighted in
Table~\ref{tab_lines_carbon}), often discussed in the literature
(e.g., \citealt{nieva08} or \citealt{daflon04}), and visible in different
temperature ranges (see Figs.~\ref{HD36512} to ~\ref{CYGOB2-7}). 

In each panel of Fig.~\ref{d35_cprof}, the black profiles refer to
model D35. In the upper panels, red profiles correspond to the same
model, however with \Teff\ increased by 1.5~kK, while the green
profiles, in turn, correspond to a \Teff\ reduced by 1.5~kK. Thus, we
are able to study the variation of important carbon lines within the
typical uncertainty of \Teff. Moreover, the upper panels also display
profiles colored in blue, which also correspond to model D35, but now
with a mass-loss rate (\mdot) reduced by a factor of three, to
estimate the impact of variations in this parameter. The effect of
this reduction becomes most obvious for supergiant models (as for
example displayed in Fig.~\ref{s35_cprof}). 

In the lower panels, we study the reaction to variations in \logg.
Here, the red lines display profiles for models with 
\logg\ increased by 0.2~dex, while the green lines correspond to 
an analogous reduction.

As shown in the upper left panel, the decrease of temperature leads to
a deeper \CII\ absorption (green line), while rising $\Teff$ results
in a shallower \CII\ profile (red line). This effect is easily
understood: lower temperatures increase the fraction of low ionized
stages, while higher temperatures favor the presence of higher ions,
in this case \CIV. From the lower left panel, we see that for the
\CII\ profile a decrease of \logg\ (green line) leads to a shallower
line (less recombination), while the opposite is seen once \logg\
increases (red line). 

The panels on the right present the reversed effects for \CIV, as
expected. For \CIII\,5696 (middle panel), on the other hand, the behaviour is quite
complex, and has been explored comprehensively by
\citet{MartinsHillier12}. Briefly, the strength of \CIII\,5696 depends
critically on the UV \CIII\ lines at 386, 574, and 884\,\AA, because
these lines control the population of the lower and upper levels of
specific optical \CIII\ lines including \CIII\,5696. Indeed, we find a
very sensitive reaction of this line on small variations in $\Teff$
(upper middle panel), and a similar effect when varying \logg\ (lower
middle panel). 
Without going into further details, during our tests we were able to
reproduce all basic effects described by \citet{MartinsHillier12},
both regarding \CIII\,5696, and also for the triplet \CIII\,4647-50-51.

The consequences of a reduction in \mdot\ are clearly seen in the
corresponding ionization fractions, where our D35 model with lower 
\mdot\ displays less \CII\ and \CIII\ (less recombination) in the wind
($\sim$ 1~dex). Though these differences do not affect the line
profiles in a notable way, the weak effect seen in the middle and
right panel indicates that these lines are not completely
photospheric.

Fig.~\ref{s35_cprof} displays a similar study on the reaction of
specific carbon lines, now for the supergiant model S35. Since
for supergiants the \CII\ lines are already very weak or absent in
this temperature range, we display another important \CIII\ line
instead of the \CII\ profile. Indeed, 
\CIII\,4068-70 behave similarly to what has been discussed for 
the \CII\ line in the previous figure. \CIII\,5696 (the one 
with complex formation!) is
now in emission, for all cases shown, and \CIV\,5801 starts to
display a P-Cygni shape. For this specific model, a reduction of
\logg\ by 0.2~dex brings the model very close to the Eddington limit
($\Gamma_{\rm e} \approx 0.77$ already for a pure Thomson scattering
opacity). The corresponding stratification becomes very uncertain, and
we refrain from displaying corresponding profiles.

Since a supergiant (model) exhibits a denser wind than a dwarf, the
effects of a mass-loss reduction on the line profiles are more obvious
than for model D35. Also here, the model with reduced \mdot\ displays
a lower fraction of \CII\ and \CIII\ in the outer wind. Particularly 
in the line forming region, however, the ionization fractions of all
ions become {\it larger}. The leftmost panel shows that a reduction of
\mdot\ leads to a stronger \CIII\,4068-70 absorption, where the effect is
even more pronounced than the effect of the temperature reduction or
the gravity increase. In the middle panel, the effect is similar, now
acting on an emission profile. Again, we see a larger response than on
the temperature decrease, which is also true for the right panel.
Additionally, the P-Cygni shape almost vanishes, due to the inward shift of
the line-forming region.

Finally, and for completeness, Appendix~\ref{app2} provides the same
analysis, now for the cooler and hotter dwarf/supergiant models at
30~kK and 40~kK, and for partly lower changes in \Teff\ and \logg. In
addition to mostly similar reactions as described above, we note
the different reaction of \CIII\,5696 on the variation of \logg\ in the
supergiant models: While for largest \logg\ (red lines) both S30 and
S35 yield the lowest emission, this behaviour switches for S40, where
the highest \logg\ results in the largest emission. Again, this
non-monotonic behaviour is due to the complex formation process of
this line.

Overall, the tests performed in the section indicate that most of our
strategic carbon lines are quite sensitive to comparatively small
variations of the stellar parameters, variations that are within the
precision of typical atmospheric analysis of massive stars performed
by means of H and He. Moreover, some of them depend on UV transitions
(as, e.g., \CIII\,5696).  Since X-ray emission affects UV lines, we
need to check which of our optical lines indirectly depend on the
strength of the X-ray emission (see Sect.~\ref{results_xrays}).

We conclude by the end of this test that even in those cases where the
stellar parameters are ``known'' from a H/He analysis, a small model grid
needs to be calculated for each stellar spectrum which should be
analyzed with respect to carbon. This grid needs to be 
centered at the (previously) derived \Teff\ and \logg\ values found
from  H/He alone, and should extend these values in the ranges
considered above\footnote{We note that a variation of the
wind-strength might be required as well.}. 
One of these models should then allow 
for a plausible fit for the majority of our \cions\ lines (and not
destroy the H/He fit quality), for a unique abundance and
micro-turbulent velocity, \vturb. 

Finally, we emphasize that all the tests discussed thus far only give
a first impression on the capabilities of our new model atom. The
quality and reliability of these results can be estimated only via a
detailed comparison with observations, for a large range of stellar
parameters. A first step into this direction is the main topic of the next
section.

\section{A first comparison with observed carbon spectra}
\label{results}

After having thoroughly tested our new model atom as far as it is
possible without relying on reality, we will now take a first step 
towards a comparison with observations. This will be done using the
philosophy just outlined above. At the end of this section, we study
the impact of X-rays on the optical carbon lines as a preview for a
future analysis. 


\subsection{Basic considerations}

\renewcommand{\arraystretch}{1.2}
\begin{table*}
\center \caption{Stellar and wind parameters adopted for and derived from
fitting the H/He plus carbon lines displayed in Figs.~\ref{HD36512}
to~\ref{CYGOB2-7}. See text.}
\label{tab_spectra}
\begin{tabular}{lllcccccccl}
\hline 
\hline
Name &  SpType-LC &  Instrument                & S/N      &  \vsini       & \vmac         & Teff
& log $g$ & $Y_{\rm He}$ & $\log Q$ & [C/H] \\
     &            &     & at 4500\,\AA    & {\tiny{(\kms)}} &{\tiny{(\kms)}}  & (kK) & [dex]   &    & [dex]  & [dex]  \\
\hline
HD\,36512  & O9.7V & HERMES & 210 & 13 & 33 & 33.8 & 4.02 & 0.105 &$-13.4$&     8.25 $\pm$ 0.22 \\ 
HD\,303311 & O6V   & FEROS  & 148 & 47 & 61 & 41.2 & 4.01 & 0.107 &$-13.0$&     8.33 $\pm$ 0.25 \\
HD\,93128  & O3.5V & FEROS  & 186 & 58 & 56 & 48.8 & 4.09 & 0.103 &$-12.7$&     8.23 $\pm$ 0.30  \\
HD\,188209 & O9.5Iab/I & HERMES & 207 & 54 & 93 & 30.3 & 3.03 & 0.145&$-12.4$ & 8.23 $\pm$ 0.25 \\
HD\,169582 & O6Ia  & FEROS & 71 & 66 & 97 & 39.0 & 3.70 & 0.225 &$-12.3$ & 8.33 $\pm$ 0.20  \\
CygOB2-7   & O3I   & FIES  & 31 & 75 & 10 & 51.0 & 4.09 & 0.139 &$-12.1$ & 8.03$^{+0.3}_{-0.4}$ \\
\hline
 \end{tabular}
\end{table*}
\renewcommand{\arraystretch}{1.0}

For our comparison with observations and a first analysis, we used six
spectra of presumably single O-type stars in different temperature
ranges, all of them relatively slow rotators. 

The reduced and normalized spectra were kindly provided and extracted
from the work by Holgado et al. (2017, submitted to A\&A).
In this work, the parameters of a large sample of Galactic O-stars
were obtained by quantitative H/He spectroscopy using {\sc FASTWIND},
where we have already summarized some observational details in
Sect.~\ref{obs_data}.

For our sub-sample, we double-checked their results by an independent
{\sc FASTWIND} analysis (fitting by-eye, contrasted to the
semi-automatic fitting method applied by Holgado et al. using
pre-calculated grids of synthetic spectra and the GBAT-tool,
\citealt{simon-diaz11b}), and found values agreeing on a 1-$\sigma$
level. We also checked the radial velocities using the H/He lines, and
confirmed the values provided by Holgado et al.  for almost all stars
(differences less than 10~\kms), except for CygOB2-7, where we found a
difference of 20~\kms.

As shown in Sect.~\ref{para_dep}, the error bars on the stellar/wind
parameters derived from H/He alone are quite large when accounting for
the sensitivity of the carbon lines. Therefore, after having defined a
first guess of these parameters, there is still a sufficiently large
interval in \Teff\ and \logg\ to vary those parameters and to find 
the best matching carbon ionization balance (in those cases where more
than one ion is present), while preserving the overall fit-quality of
the H/He lines. To this end, we varied \Teff\ and \logg\ inside
intervals of  $\pm$1,000~K
and $\pm$0.1~dex, respectively, centered at the initial values derived
by Holgado et al..

These authors also estimated the wind-strength parameter, $Q =
\mdot/(\Rstar \vinf)^{1.5}$ (e.g., \citealt{puls05}) for each star in
their sample, though they did not provide individual values for \mdot,
\Rstar, and \vinf\ as required for the {\sc FASTWIND} input. We
estimated these quantities using their $Q$-values, an estimate of
\vinf\ (via \vesc, using \logg, \Rstar, and \Teff, see
\citealt{KP00}), and an adopted stellar radius, \Rstar, prototypical
for the considered spectral type. 

Because of the weakness of most lines and the blending problem,
rotational broadening is of major concern for a meaningful comparison
of synthetic and observed spectra. Usually, hot massive stars
are fast rotators (e.g., \citealt{Herrero14}), and any
large value of \vsini\ (particularly in combination with a significant
extra-broadening due to ``macroturbulence'', \vmac)
makes the majority of carbon lines very shallow or even too shallow to
be identified. Thus, we restricted our sub-sample to comparatively slow
rotators, and double-checked also the \vsini\ and \vmac\ values
derived by Holgado et al. (indeed, we found very similar results). 
Table~\ref{tab_spectra} summarizes the final values derived from our
fits to the optical H/He\footnote{including \Ha\ and
\HeII\,4686} {\it and} C-lines, for all objects considered.

As we have double-checked all stellar and wind-parameters (but varied
\Teff\ and \logg\ to improve on the carbon ionization balance), and
these parameters turned out to be sufficient to reproduce the H/He and C
profiles, we have not performed an independent error analysis, and
refer to the values suggested by Holgado et al.. On the other hand,
since we additionally derive the carbon abundances, we need to
estimate their uncertainties. This was done by using the finally
derived values, and then calculating two more models with a carbon
abundance varied by $\pm$0.2~dex (or more, for specific models).
This allowed us to obtain a rough estimate on the associated
uncertainty, as displayed in Table~\ref{tab_spectra} and
Figs.~\ref{HD36512} to~\ref{CYGOB2-7}.

Our sub-sample comprises three dwarf and three supergiant O stars,
observed with different instruments. However, all spectra cover the
wavelength range relevant for this work. As in our previous tests, we
cover the same interval of temperature, from $\sim$30~kK to
$\sim$50~kK. Thus, we expect to analyze \CII/\CIII\ for the coldest
stars, while the hotter stars provide an opportunity to check our
precision in reproducing \CIII\ and \CIV\ lines. 

In the following, each of the spectra and corresponding fits will be 
discussed in fair detail. Figs.~\ref{HD36512} to~\ref{CYGOB2-7} 
present the observed spectra and our best solution (in black),
corresponding to the parameters as given in Table~\ref{tab_spectra}.
The red and the blue lines refer to a carbon abundance increased and
decreased by 0.2~dex, respectively. These profiles provide us with an
estimate on the error of our finally derived abundance (see above),
and also allow us to identify which of the lines are more or less
sensitive to abundance variations.

We remind here that lines sensitive to significant X-ray emission from
wind embedded shocks (discussed in Sect.~\ref{results_xrays}) have
been marked as ``X-ray dependent'' in the last column of 
Table~\ref{tab_lines_carbon}. 


\subsection{Details on individual spectra}
\label{stellar_sample} 

\begin{figure*}
\center
{\includegraphics[width=160mm]{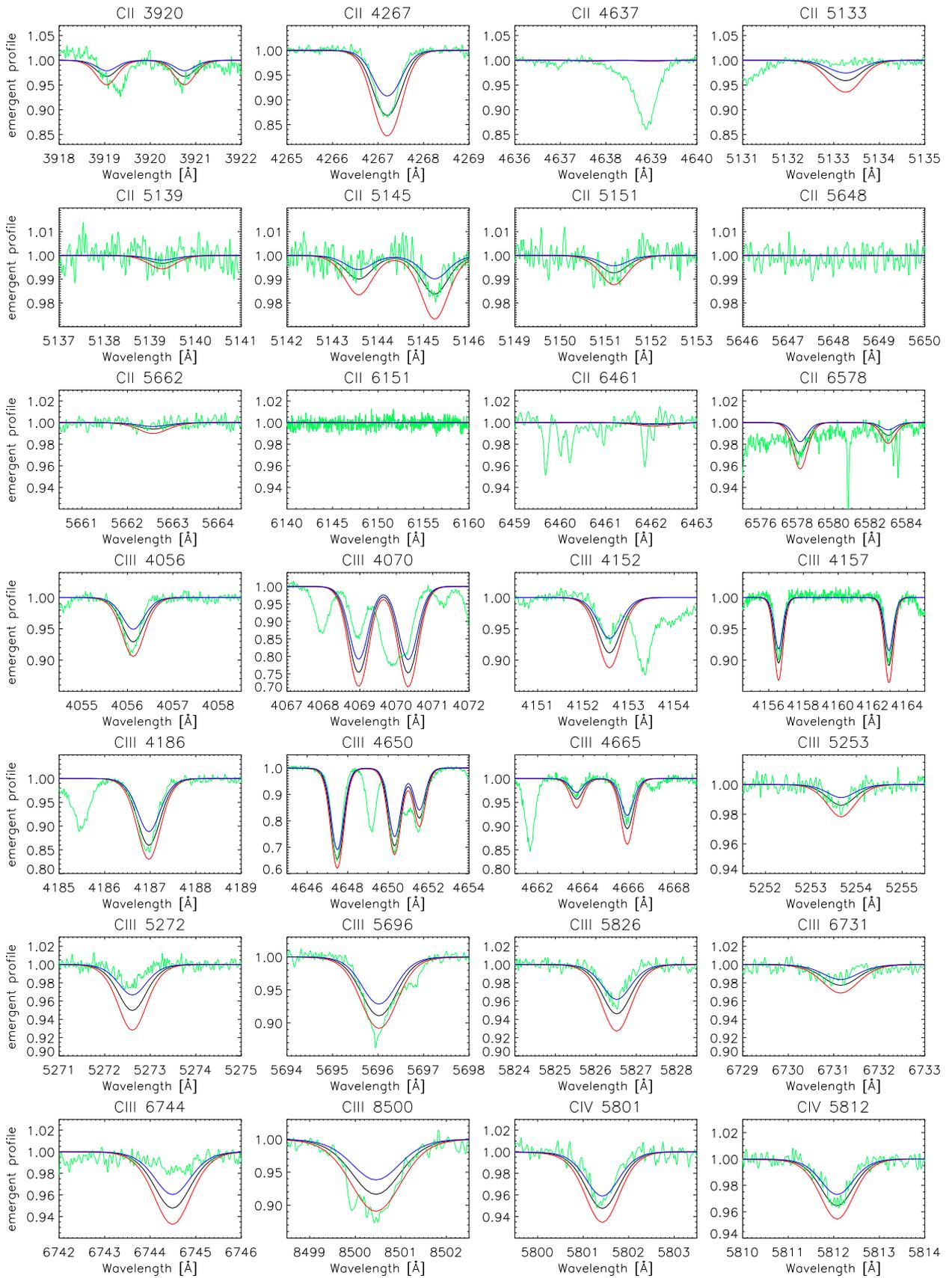}}
\caption{Observed carbon spectrum of HD\,36512 (O9.7V, green), and synthetic
lines (black), calculated with [C/H] = 8.25~dex. 
The red and blue profiles have been calculated with an abundance
increased and decreased by 0.2~dex, respectively.}
\label{HD36512}
\end{figure*}

{\it HD\,36512} ($\upsilon$ Ori) is an O9.7V slow rotator, observed
with the HERMES spectrograph (see Fig.~\ref{HD36512}). We fitted the
H/He and C lines with a temperature of 33.8~kK and \logg\ = 4.02. The
obtained stellar parameters agree well with the values derived by
Holgado et al. This is one of the stars where {\it all} the carbon
ions have well-defined observable lines.

Our synthetic spectra reproduce quite well the \CII\ and \CIV\ lines.
\CII\,4637 is absent (\OII\,4638.9 dominates the range), as well as
\CII\,5133. For \CIII, basically all lines are reproduced, except for
\CIII\,6744, \CIII\,5272, and the \CIII\,4068-70 doublet, which always
seems to
indicate a lower carbon abundance than inferred from the other lines,
and in parallel is strongly influenced by \OII\,4069.8. At least for
this object, the discrepancy seems to be stronger for the \CIII\,4068
component than for its $\lambda$4070\,\AA\, companion.

We derive a carbon abundance of [C/H] = log C/H + 12 =
8.25~dex\footnote{(i) in terms of number density. (ii) for reference,
the solar carbon abundance is 8.43$\pm$0.15~dex according to
\citet{asplund09}, while \citet{Przyb08} estimated, from quantitative
spectroscopy of B-dwarfs, a carbon abundance of 8.32$\pm$0.03~dex as a
cosmic abundance standard for the solar neighborhood.}, which brings
most carbon lines into agreement. Few of our lines point to slightly
different abundances (e.g., \CIII\,5696-6744-8500), and therefore we
estimate a range of $\pm$0.22~dex for the involved uncertainties. This
spectrum/object is an example for an ideal scenario, mainly due to the
low rotation rate (\vsini\ = 13~\kms) and low macroturbulence (\vmac=
33~\kms), where our carbon model produces very satisfactory results.  
\citet{martins15a} have analyzed this star as well, and they derived,
in addition to rather similar stellar parameters, also a carbon
abundance ([C/H] = 8.38 $\pm$ 0.15) that is consistent with our result.

\begin{figure*}
\center
{\includegraphics[scale=0.87]{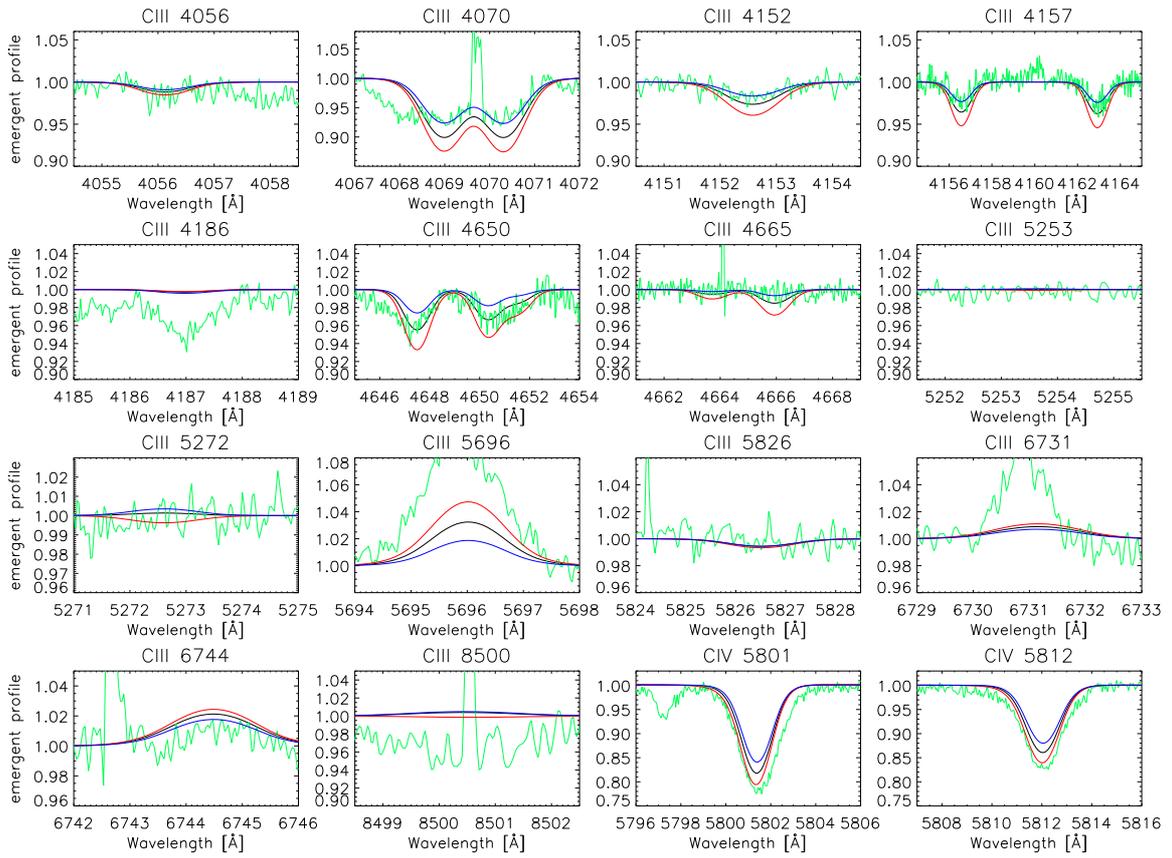}}
\caption{As Fig.~\ref{HD36512}, but for HD\,303311 (O6V), and a carbon
abundance of 8.33~dex. The optical \CII\ lines are not visible, and
thus not displayed.}
\label{HD303311}
\end{figure*}

{\it HD\,303311} is an O6V star with 47~\kms\ projected rotational
velocity, and a macroturbulence of 61~\kms\ (Fig.~\ref{HD303311}). The
spectrum has been collected with the FEROS spectrograph. We obtained a
final value of 41.2~kK for the temperature and of 4.01 for \logg, both
slightly adjusted after the reproduction of the H/He lines to the best
agreement with the different carbon lines. At this temperature (and
rotational velocity), the lines of \CII\ already vanish, and the
\CIII\ profiles are weak, while the \CIV\ lines are still easily
detectable. Our synthetic lines show a good reproduction of the 
\CIII\ lines. Once more, \CIII\,4068-70 indicate a lower abundance
when compared to the other \CIII\ profiles, however the difference is
not larger than 0.2~dex. \CIII\,6731 surprisingly displays an
emission profile. There seems to be a disagreement between the carbon
abundance indicated by the \CIII\ and \CIV\ lines. Both \CIV\ profiles
point to a higher [C/H]-value, but again the difference is not
larger than 0.22~dex. The best compromise was found for a carbon
abundance of 8.33$\pm$0.25~dex. 

\begin{figure*}[b]
\center
{\includegraphics[scale=0.87]{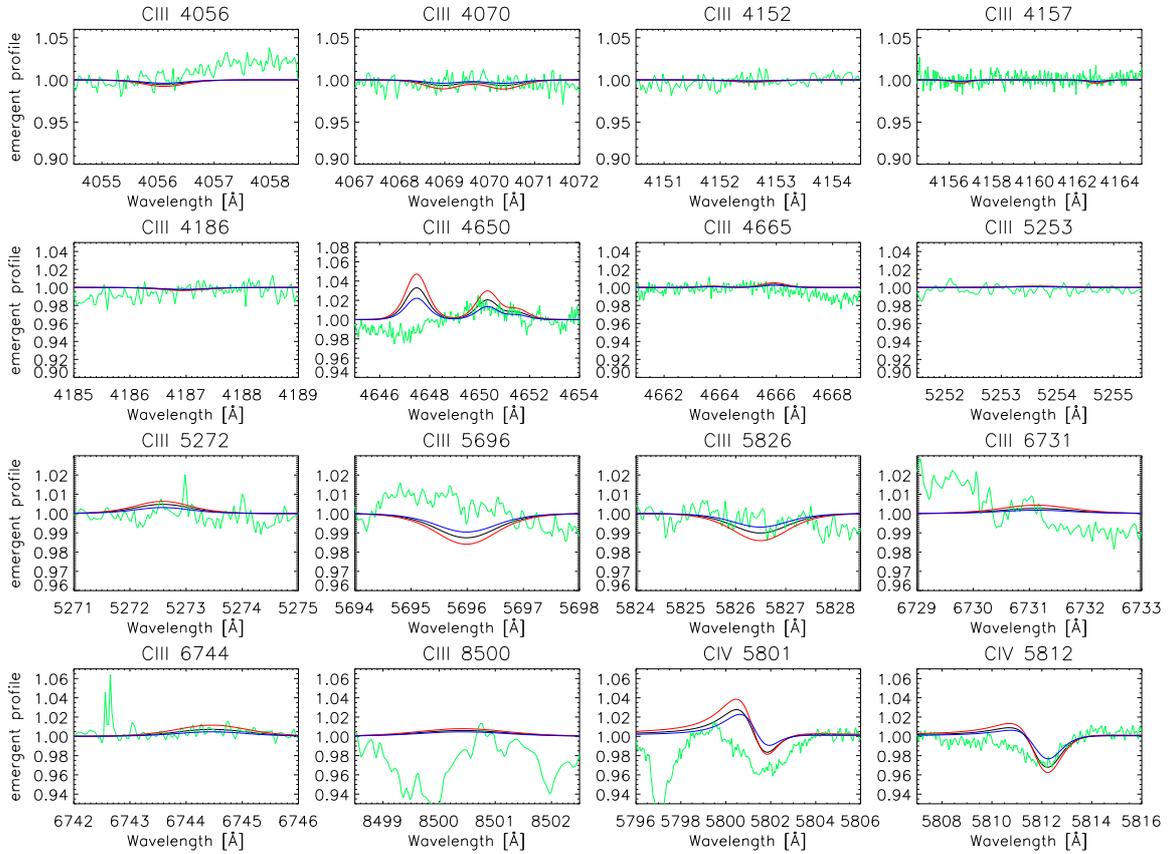}}
\caption{As Fig.~\ref{HD303311}, but for HD\,93128 (O3.5V), and 
a carbon abundance of [C/H] = 8.23~dex.} 
\label{HD93128}
\end{figure*}

{\it HD\,93128} is an O3.5V star rotating with 58~\kms, a
macroturbulence of 56~\kms, and was observed with the FEROS
spectrograph (Fig.~\ref{HD93128}).  The temperature has been decreased
by 300~K from the value obtained from the pure H/He analysis, but is
still in agreement with the value from Holgado et al. when considering
their 1-$\sigma$ interval. We used 48.8~kK for the temperature, and
4.09 for $\logg$. In this temperature regime, some weak signs of
\CIII\ might be seen only by chance. Furthermore, also
the \CIV-analysis becomes complicated, because the lines start to switch
from absorption to emission, and a distinction 
from the continuum is difficult in this case.

Anyhow, at least a rough estimate for the carbon abundance might be
provided, both from \CIII\ and \CIV. The black line in
Fig.~\ref{HD93128} fits the weak sign of \CIII\,4650, and also
\CIV\,5812, and we infer [C/H] $\approx$ 8.23~dex. Due to the
very low number of available lines, we adopt a larger uncertainty in
our estimate, $\pm$ 0.3~dex.

\begin{figure*}
\center
{\includegraphics[width=160mm]{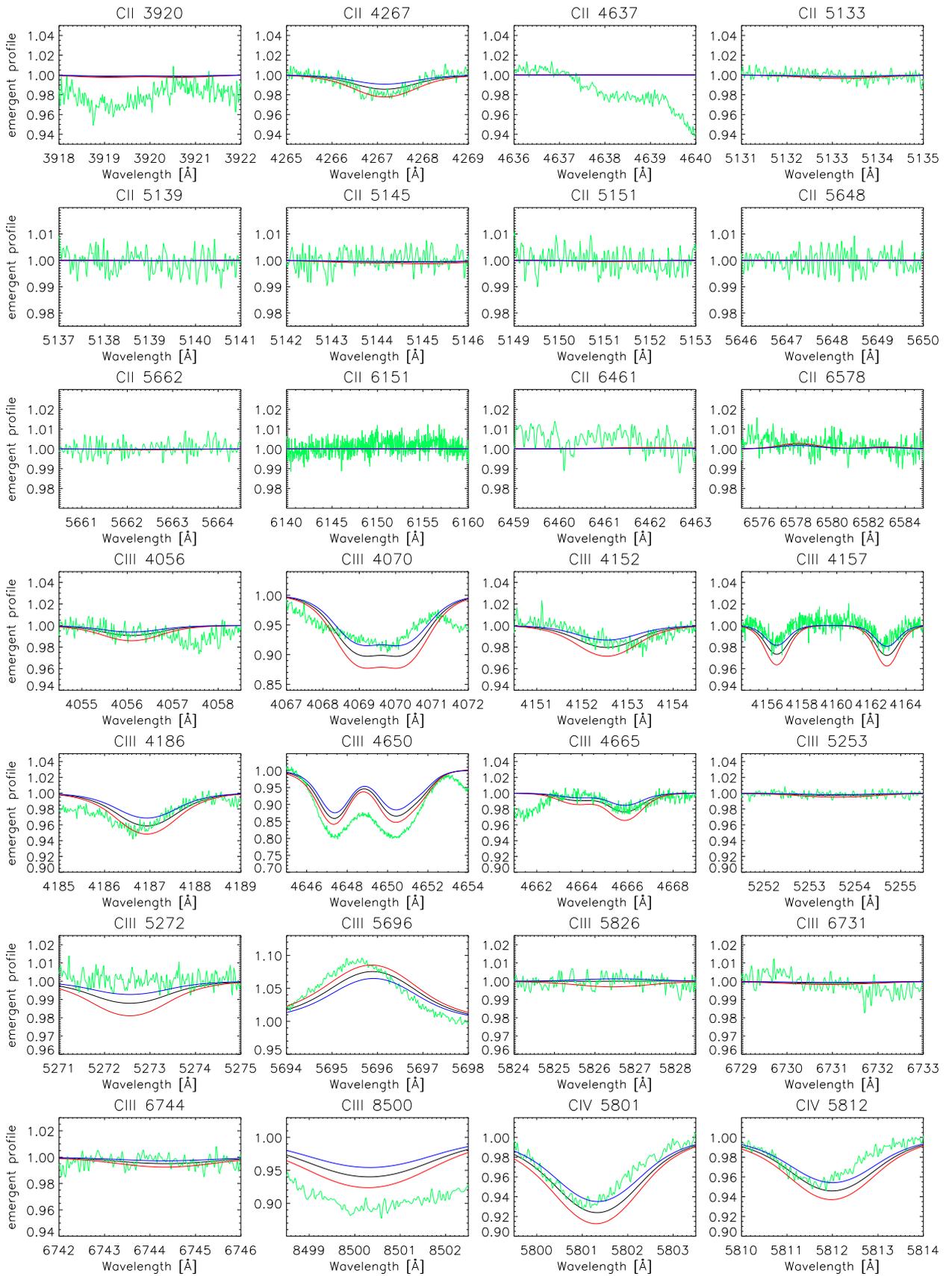}}
\caption{As Fig.~\ref{HD36512}, but for HD\,188209 (O9.5Iab), and a 
carbon abundance of 8.23~dex.} 
\label{HD188209}
\end{figure*}

{\it HD\,188209} is an O9.5Iab star with \vsini\ of 54~\kms, a
macroturbulence of 93~\kms, and has been observed with the HERMES
spectrograph (Fig.~\ref{HD188209}). The temperature and gravity
obtained from fitting the H/He lines agree with the stellar parameters
derived from Holgado et al., and were also used in our final model
including the carbon line diagnostics (\Teff = 30.3~kK, \logg\ =
3.03). \CIII\ and \CIV\ lines are easily identified, while \CII\ lines
are not present in this case, except a subtle sign of \CII\,4267,
which is well reproduced by our synthetic profile. The \CIII\ and
\CIV\ lines, even being weak, are well described by the synthetic
profiles, and the discrepancy of \CIII\,4068-70 is somewhat lower than
found in the cases above. Here, \CIII\,4650 shows the largest
deviations. Our final solution for [C/H] is 8.23~dex, and due to
non-fitting lines we increase our error budget to $\pm$0.25~dex. Also
this star has been analyzed by \citet{martins15a}. Again, the stellar
parameters are in very good agreement, but here the derived carbon
abundance ([C/H] = 7.85 $\pm$ 0.3~dex) only marginally overlaps with our
value within the quoted error intervals. 

\begin{figure*}
\center
{\includegraphics[scale=0.87]{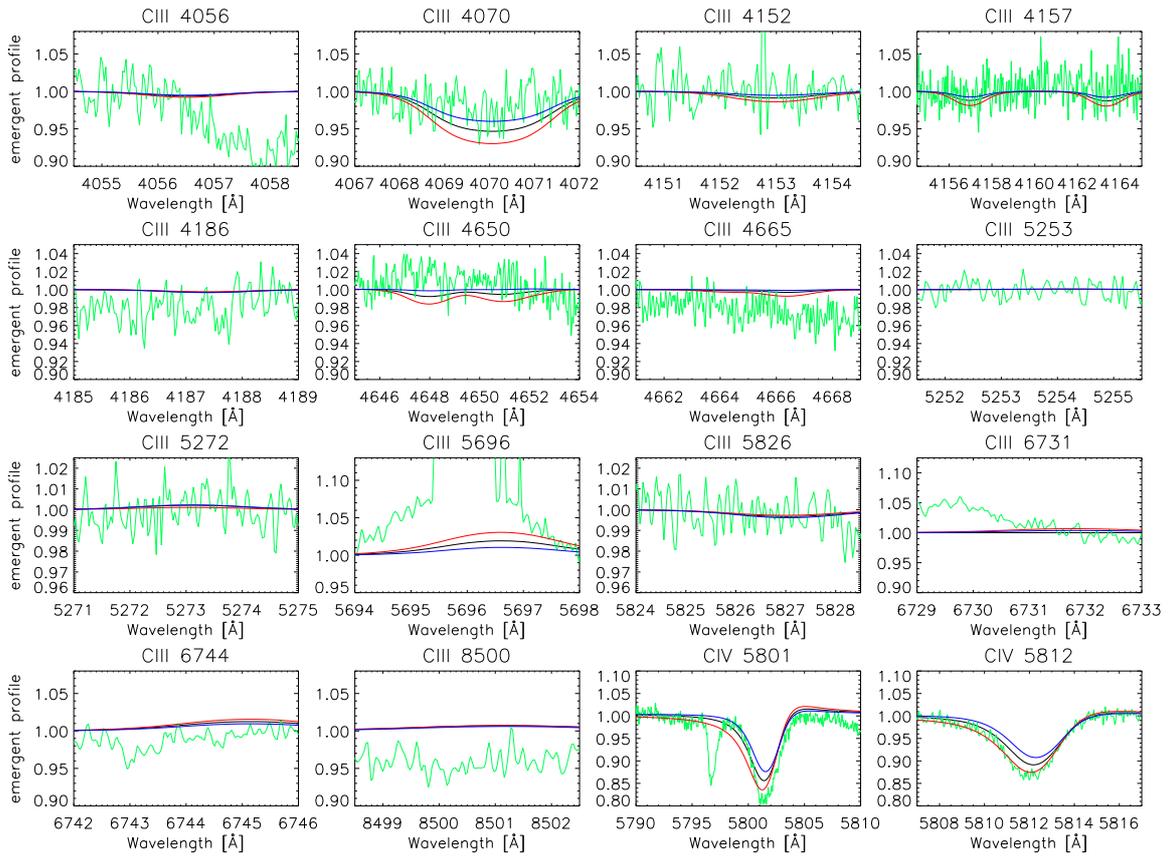}}
\caption{As Fig.~\ref{HD303311}, but for HD\,169582 (O6Ia), and carbon
abundance of 8.33~dex.} \label{HD169582}
\end{figure*}

{\it HD\,169582} (O6Ia) rotates with \vsini\ = 66~\kms, has a
macroturbulence of 97~\kms, and was observed with the FEROS
spectrograph (Fig.~\ref{HD169582}). A temperature of 39~kK and \logg\
of 3.7 were used to synthesize the carbon lines. Both values agree
with the ones suggested by Holgado et al.. \CIII\ is very weak and
almost invisible, and only the \CIV\ profiles are easily visible. Firm
conclusions about \CIII\ are not possible, though we note that the
synthetic lines indicate a weak signal. 
A carbon abundance of 8.33~dex gives a fair compromise for the
\CIII/\CIV\ lines, though \CIV\ seems to indicate a slightly higher abundance
than \CIII. We note however that none of the lines requires an
abundance outside the $\pm$0.2~dex interval.

\begin{figure*}
\center
{\includegraphics[scale=0.87]{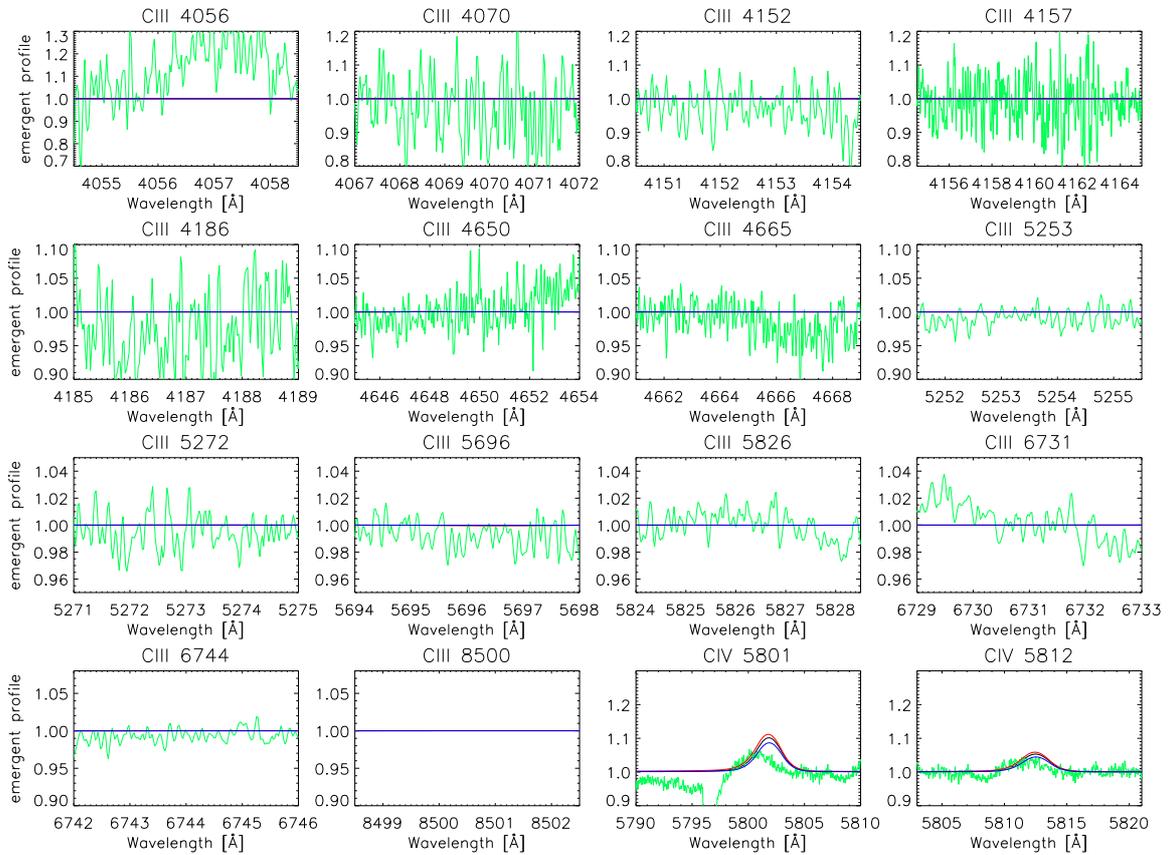}}
\caption{As Fig.~\ref{HD303311}, but for CygOB2-7 (O3I), and a 
carbon abundance of 8.03~dex.} 
\label{CYGOB2-7}
\end{figure*}

{\it CygOB2-7} is one of the few O3I stars in the Milky Way. Its
spectrum (Fig.~\ref{CYGOB2-7}) has been recorded by the
FIES-spectrograph, and extends ``only'' to a maximum of 7000\,\AA, so
that \CIII\,8500 is not available. We note that this spectrum has
the lowest S/N within our sub-sample. A \Teff\ of 51~kK and a \logg\
of 4.09 (together with $\vsini\ = 75~\kms$ and $\vmac\ = 10~\kms$(!)) enable
a satisfactory fit to the H/He lines.  In this temperature regime,
only \CIV\ is visible, switching from absorption to emission (at least
at the given \mdot). This behaviour complicates the reproduction of
the \CIV\ profiles, and forbids any stringent conclusions. Especially
in this case, one would also need to analyze the UV spectrum. If we
believe in the ionization equilibrium and the mass-loss rate, we
derive an abundance around [C/H] $\approx$ 8.0, which would be the
lowest value in our sample. From the fit quality and since we have to
firmly rely on our theoretical models (no constraint on the ionization
equilibrium), we adopt an asymmetric error interval, $-0.4$ and $+0.3$~dex.

As mentioned in Sect.~\ref{carbon_lines}, one of the ``classical''
problems in carbon spectroscopy is an inconsistent abundance implied
by \CII\,4267 and \CII\,6578-82. These lines are clearly visible and 
well reproduced with the same value of [C/H] in our coldest dwarf,
HD\,36512. This provides strong evidence that our present data are
sufficient to overcome this issue. Also for our coldest supergiant, 
HD\,188209, \CII\,4267 is present and well reproduced. On the
other hand, \CII\,6578-82 is absent, and thus no further
conclusions can be asserted.

We finish this section by noting that part of the problems in
fitting certain lines might be related to our assumption of a smooth
wind. Effects due to clumping will be investigated in a forthcoming
paper.

\clearpage

\clearpage

\subsection{Which lines to use?}
\label{set_lines} 

After our first analysis, we acquired enough experience to judge in
which lines to ``trust'' when deriving carbon abundances. In
Table~\ref{tab_lines_carbon}, we provided a comprehensive list,
comprising many more lines than previously studied, which are strong
enough to be easily identified in different temperature ranges.
Instead of describing which of these lines are the most useful, we
summarize which may be discarded, since this results in a shorter list.

For \CII, the range around \CII\,4637 is dominated by 
\OII\,4638, and therefore the carbon lines are barely visible. 
\CII\,5648-62 are isolated lines which can be important, but are
not visible in the range of spectral types studied in this work
(O9-O3). The same is true for \CII\,6461. The lines at 5139 and 
6151\,\AA\ are formed by transitions with low oscillator strength, 
and might be too weak for a meaningful spectral diagnostics. Excluding 
these lines, we were able to identify all the other \CII\ lines as 
listed in Table~\ref{tab_lines_carbon} in the observed spectra 
(for the cooler spectral types), and to use them within our analysis.

\CIII\ provides us with the largest number of lines, when considering
the {\it complete} O-star range. Particularly, all the listed lines
are visible in the coldest dwarf of our sample (Fig.~\ref{HD36512}). 
\CIII\,4068-70 always (i.e., for the complete temperature range) point
to lower abundances (compared to the majority of other lines), and it
might be that particularly the $\lambda$4068\,\AA\ component is either
mistreated by our approach, or that there is a problem with its
oscillator strength.  The lines at 4650 and 5696\,\AA\ always deserve
special attention, because of their complicated formation process,
even though we were able to reproduce these lines well in the majority
of cases studied here. The lines at 5826, 6731, and 6744\,\AA\ are
also good diagnostics, but vanish quickly for spectral types earlier
than O9.

For \CIV, basically four lines are available in the optical range, 
but the ones at 5016-18\,\AA\ are outshone by \HeI\,5015.
Therefore, and to our knowledge, all optical \CIV\ analysis performed
until to-date have concentrated on \CIV\,5801-12, and this most likely 
will not change in future.

Discarding the lines quoted above, we end up with a list 
of 27 lines from \cions\ that are useful for determining 
reliable carbon abundances, indicated in bold-face in 
Table~\ref{tab_lines_carbon}.

\subsection{Impact of X-rays}
\label{results_xrays}

In a previous paper \citep{carneiro16}, we already discussed the impact of
X-ray radiation on the ionization stratification of different ions,
including carbon. Here we investigate which of the {\it optical} lines are
affected by emission from wind-embedded shocks, and how intense the X-ray
radiation must be to have a relevant impact on the lines. As pointed out
before, purely photospheric lines without any connection to UV-transitions
should not be affected by X-rays, at least in principle. However, lines that
are purely photospheric for thin winds are partly formed in the wind when
the mass-loss rate becomes larger, and also the lower boundary of the X-ray
emitting volume is important in controlling how much X-ray/EUV radiation can
reach the photosphere. Even more, since the X-ray luminosity scales with the
mass-loss rate (or, equivalently, with the stellar luminosity, e.g.,
\citealt{OwockiSundqvist13}), carbon lines in high-luminosity objects might
become affected by X-ray emission even when they are not connected with
UV-transitions.

The main idea of our study is to adopt the strongest possible (and
plausible) shock radiation, and to check which lines will change. For
the present analysis, few parameters will describe the shock radiation
in each model, leaving the others at their default (see
\citealt{carneiro16} for details). These are the X-ray filling factor,
$f_{x}$, which is related to (but not the same as) the (volume)
fraction of X-ray emitting material, and the maximum shock
temperature, \Tshockmax. Both are set here to the maximum values used
in our previous analysis: $f_{x}$= 0.05, and \Tshockmax= 5$\cdot
10^6$~K. Besides this ``maximum-model'', we checked also the impact
for intermediate values of the X-rays parameters ($f_{x}$= 0.03, and
\Tshockmax= 3$\cdot 10^6$~K). Another important parameter is the onset
of X-ray emission, \Rmin. Guided by theoretical models on the
line-instability and/or by constraints from X-ray line diagnostics, 
\Rmin\ is conventionally adopted as $\sim$1.5~\Rstar\ (e.g.,
\citealt{hillier93}, \citealt{Feldmeier97b}, \citealt{Cohen14}). Since
we want to maximize any possible effect from the X-ray radiation, we
set \Rmin = 1.2~\Rstar.

Before turning to the general results of our simulation, we remind on
the sensitivity of \CIII~5696 and \CIII~4647-50-51, showing
significant changes in strength and shape for small variations of
local conditions in the 30-40~kK regime (see Fig.~\ref{d35_cprof} and
\citealt{martins12} for a thorough analysis). As expected (both
transitions are connected to UV resonance lines), these lines are
indeed sensitive to the presence of X-rays.

After checking all lines tabulated in Table~\ref{tab_lines_carbon}
regarding a potential influence of X-ray emission, no changes were
found for the 30~kK and 35~kK dwarf and supergiant models. Even for
\CII\ in these coolest models, no impact was seen, which indicates that
either the X-ray radiation is still to weak (because of low
mass-loss rates), or that it cannot reach the photosphere.

From 40~kK on, however, the situation changes. In almost all cases,
only the \CIV\ lines become weaker, and by a considerable amount for
supergiants (see below) and our D50 model. Most \CIII\ lines become
only marginally stronger or weaker, if at all, and the only more
significant reaction is found for the ``complicated''
\CIII\,4647-50-51 and \CIII\,5696 lines. When including shock
radiation, their strength increases at hottest temperature(s),
comparable to an increase in carbon abundance of 0.1~dex, 

Beyond 40~kK, the ionization fraction of \CIV\ decreases (both in the
line-forming region and the wind) when the X-ray emission is included.
For dwarfs, the corresponding line-strengths of \CIV\,5801-5811 (in
emission) decrease in parallel, by an amount still weaker than 0.1~dex
in [C/H].

For the supergiants, this effect becomes stronger in the 40 to 45~kK
regime, while for S50, finally, the impact of X-rays on
the \CIV\ lines becomes weak again, presumably because in this
temperature range the {\it stellar} radiation field dominates in
controlling the ionization equilibrium. Note that for the D50 dwarf
model the changes remain considerable though.

\begin{figure}[t]
\resizebox{\hsize}{!}
{\includegraphics[angle=90]{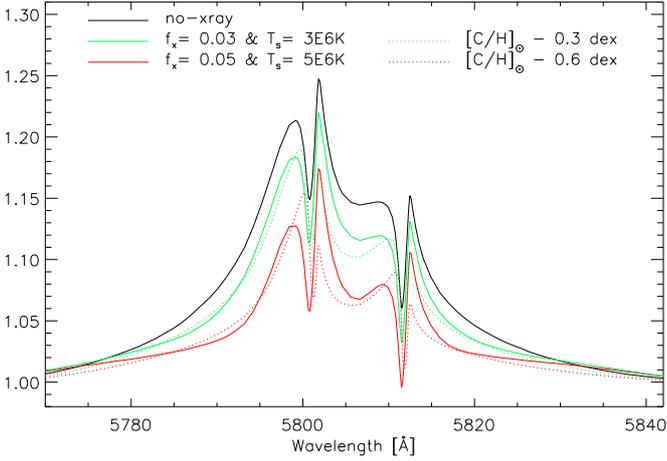}}
%
\caption{Impact of X-ray radiation on the \CIV~5801-12 lines, for a
supergiant model with \Teff\ = 40~kK (S40). Black: Line profiles for
the ``standard'' model without shocks; green: model with an
intermediate X-ray strength; red: model with strong X-ray radiation.
The dotted profiles refer to alternative, no-shock models with reduced
carbon abundance, mimicking the effect of shock radiation. We note
that the displayed sensitivity is strongest for the considered lines
and parameter range, see text.}
\label{plot_xrays}
\end{figure}

In Fig.~\ref{plot_xrays}, we detail this behaviour, for our S40 model,
where the effect is strongest.  The black line represents a model
without X-rays, the green continuous line corresponds to a model with
intermediate shock radiation, and the red continuous line displays the
model with our strongest X-ray emission. The dotted profiles give an
impression of a corresponding decrease in carbon abundance which would
be necessary to mimic the X-ray effect, which is 0.3 and 0.6~dex,
respectively. The other way round, for stars that have been analyzed
without X-rays but exhibit a strong X-ray radiation field, the
originally derived carbon abundance might need to be {\it increased}
by such an amount to compensate for the missing X-ray field. Our
investigation clearly indicates that X-rays may be important for the
\CIV\ analysis of supergiant stars with temperatures around 40 to
45~kK (e.g., the prototypical $\zeta$ Pup) and for (very) hot dwarfs, in
particular if no lines from other carbon ions are present.

Summarizing,  the changes are marginal for not too hot dwarfs, and  
affect only a few \CIII\ lines (the triplet at 4650, and \CIII\,4665, 
5696, 5826, 8500\,\AA) which might be used with a lower weight in
abundance analysis. 

In contrast and at least for supergiants in the range between 40 to
45~kK, various lines become substantially modified when accounting for
strong emission from wind-embedded shocks, in particular the
``complicated'' \CIII\ lines and the \CIV\ lines. The potential
differences in abundances derived from these lines ($\sim$0.1~dex from
\CIII\,4647-50-51 and \CIII\,5696, and $\sim$0.3 to 0.6~dex from
\CIV\,5801-5811) may complicate the analysis considerably, and we
conclude that the carbon analysis of supergiants earlier than O7
should include X-ray radiation using typical default values, as
already standard for {\sc CMFGEN}-modelling. We note (i) that this
problem might have also affected our analysis of HD\,169582, and (ii)
that X-rays might need to be considered in the analysis of (very) hot
dwarfs as well, due to their impact on \CIV.

\section{Summary and conclusions}
\label{summary}

In this work, we aimed at enabling a reliable carbon spectros\-co\-py
by means of our unified NLTE atmosphere code {\sc FAST\-WIND}. To this
end, we developed a new carbon model atom using high-accuracy data
from different databases.  The data adopted to describe the
radiative/collisional transitions have been cross-checked (with
similar data from alternative databases) to ensure their reliability.
We implemented an adequate number of levels following certain rules,
though additional levels might be required for the analysis of
IR-transitions in future studies. In total, we
considered 162 LS-coupled levels, more than 1000 radiative, and more
than 4000 collisional transitions. 

One major issue of this study was a rigorous test of our new model atom.
For this purpose, we calculated a sufficiently spaced grid of
atmospheric models, to investigate the carbon ionization structure, and
to compare with previous results using more simplified model atoms. 

Interestingly, dielectronic recombination (DR) turned out to be of
major importance for the description of \CII\ and \CIII\ in the wind,
for almost all temperatures\footnote{except for \Teff\ $\ga$ 50~kK}.
When DR is neglected, we find less recombination from \CIV\ to \CIII,
and consequently also less \CII. Typical differences amount to 1~dex.
Similar results were reproduced with WM-{\sc basic}, though we found
an important difference between the underlying data: The strengths of
the stabilizing transitions used in WM-{\sc basic} are typically
larger (up to a factor of $\sim$ 2) than those corresponding to the
strengths of the resonances included in the OPACITY Project data we
have used, leading to stronger recombination in WM-{\sc basic} models.
Even though this potential uncertainty will not affect the majority of
optical lines formed in the photosphere, it can be
influential in upcoming analysis of UV wind lines. In the end, we
prefered to use the OPACITY Project data, for reasons outlined in
Sect.~\ref{tests_dr}.

Unfortunately, optical carbon lines are rare, and often blended by
other lines, which hampers the spectral diagnostics, particularly for
large \vsini\ and/or \vmac. Thus, we compiled and selected a
maximum set of potential diagnostic lines, including blended
ones and also those with a complex formation mechanism controlled by
UV transitions (as decribed by \citealt{martins12}).

Since the majority of metal lines are weak, they are very sensitive to
comparatively small variations of stellar parameters, and this is
particularly true for carbon lines. A change of $\pm$1000 to 1500~K in
effective temperature, or $\pm$0.2~dex in \logg, results in
considerable changes of line strength. Additionally, some of the lines
are also sensitive to \mdot\ variations. Mainly in supergiants (and
hot dwarfs), due to their dense winds, a decrease in \mdot\ by a
factor of three produces an effect stronger than a decrease of 1500~K
in \Teff\ or an increase of 0.2~dex in \logg. Thus one needs to
precisely determine \mdot\ by reproducing, e.g., \Ha\ and \HeII\,4686,
before aiming at a carbon analysis. 

As a first application of our ``new'' capability to analyze carbon
lines by means of {\sc FASTWIND}, using our newly developed model
atom, we investigated the spectra of a sample of six O-type dwarfs and
supergiants, kindly provided to us by Holgado et al. (2017, submitted
to A\&A). We first convinced ourselves that the stellar/wind
parameters derived by Holgado et al. are reproduced by our own H and He
analysis. Thereafter, we varied \Teff\ and \logg\ inside a range where
the fit quality of the H/He lines could be preserved (roughly,
$\pm$1000~K and $\pm$0.1~dex, respectively), while fitting the carbon
lines from different ions in an optimum way, with an (almost) unique
abundance. From the reaction of the carbon lines when varying the
abundances, we obtained a rough estimate on the corresponding error,
typically in the range $\pm$0.2 to $\pm$0.3~dex. 

Only for the coolest (O9.7) dwarf, lines from
all three ionization stages are present in parallel, and our analysis
resulted in a satisfactory reproduction of the ionization balance. In
most other stars, only \CIII\ and \CIV\ are visible in parallel. 
For one object with $\Teff \approx$ 40~kK (HD\,169582), 
these \CIII/\IV\ lines disagree by $\sim$0.2~dex in [C/H], which might
be related to the influence from X-ray radiation (see below).
The hottest (O3) supergiant, CygOB2-7, displays only \CIV,
which renders the analysis quite uncertain (and there are only two
suitable \CIV\ lines in the optical!). 

Anyhow, we were able to derive carbon abundances for all the
considered objects, and found in most cases a moderate depletion
compared to the solar value by \citet{asplund09}, except for CygOB2-7,
where a larger depletion by 0.4~dex was derived (though more uncertain
than the other values). Two of our cool objects had already been
analyzed by \citet{martins15a}. For both of them we confirmed rather
similar stellar parameters, but only for one of them (a dwarf) also
the carbon abundances are quite similar, while for the other (a
supergiant), there is a only marginal overlap within the errors.  

To obtain further constraints on the reliability of our new set of
synthetic carbon lines, we have to repeat the same exercise in
particular for cooler stars, since \CII\ already disappears for \Teff
$\ga$ 30 to 35~kK, in dependence of luminosity class.

From our accumulated experience of analyzing basically all optical
carbon lines, we finally provided a list of more than 25 lines of
different strength and from different ions, which we consider as
reliable carbon abundance indicators. 

As a first step towards future work, we studied the direct and
indirect (via UV-lines) impact of X-ray emission from wind-embedded
shocks onto our sample of optical carbon lines, by simulating an X-ray
radiation field that is at the upper limit of realistic values.

For the dwarf models, we found an almost negligible impact, except for
very hot objects ($\ga$50~kK). In contrast, carbon lines from
supergiants with \Teff\ $\approx$ 35~kK already show changes due to
shock radiation, and at 40 to 45~kK the impact of X-rays is strong
enough to complicate a reliable abundance measurement. Mostly, the
\CIII\ lines become stronger (corresponding to an increase of up to
0.1~dex in [C/H]; the largest changes occur in the ``problematic''
\CIII\,5696 and \CIII\,4647-50-51), and the \CIV\ lines become weaker
(corresponding to a decrease of 0.3 to 0.6~dex in [C/H]).
Consequently, it might become more difficult
(or even impossible) to find a value of [C/H] which brings different
carbon lines into agreement, when strong X-ray emission would be
present. Thus we concluded that in the spectral regime earlier than
O7\,I, it might be necessary to include the impact of X-rays by
default (though a deeper understanding of typical X-ray parameters
might be required as well). For supergiants with \Teff\ $\ga$ 50~kK, the
X-rays lose their impact, since direct ionization due to the
stellar radiation field dominates over Auger-ionization and ionization
from the EUV-component of the shock radiation (see
\citealt{carneiro16}).

This study has been performed as a first step towards a complete CNO
analysis. Particularly the investigation of the ratio N/C vs. N/O (as
already begun by \citealt{martins15a, martins15b}) will allow us to
derive better constraints on the mixing history and chemical
enrichment in massive stars than from a pure nitrogen analysis alone.
We intend to continue such work, also by including the information
provided by UV carbon lines, and by investigating the impact of wind
inhomogeneities, which might play an important role also in the UV
range, due to porosity effects and because of affecting the ionization
balance in the wind.

\begin{acknowledgements}
L.P.C. gratefully acknowledges support from
the Brazilian Coordination for the Improvement of Higher Education
Personnel (CAPES), under grant 0964-13-1. Many thanks also to 
G.~Holgado and S.~Sim\'on-D\'iaz for providing us with the optical
spectra for six Galactic O-type stars, and to N.~Przybilla and
M.~Urbaneja for their help, material and suggestions to build the
carbon model atom. Special thanks to Keith Butler for all his
explanations regarding atomic data and corresponding databases.
\end{acknowledgements}

\bibliographystyle{aa}
\bibliography{carneiro.bib}

\clearpage
\appendix

\section{Electronic states of each carbon ion} 
\label{electronic_states}

This section provides a short description of each ion considered in
our new carbon model atom, except for \CV\ which is described by the
ground level only (see Sect.~\ref{carbon_model}). All the next three
tables have the same format: the first column displays the label of
the level, the second column the electronic configuration of that
level, and the third column presents the term designation. 
Table~\ref{levels_cii} refers to \CII, Table~\ref{levels_ciii}
to \CIII, and Table~\ref{levels_civ} to \CIV. 
\begin{table}
\tabcolsep1.7mm
\caption{\CII\ levels: label, electronic configuration, and term 
         designation.}
\label{levels_cii}
\begin{tabular}{rllrrc}
\hline
\hline
\multicolumn{1}{l}{C2\_\#}
&\multicolumn{1}{l}{Configuration}
&\multicolumn{1}{l}{Term}
&\multicolumn{1}{c}{C2\_\#}
&\multicolumn{1}{c}{Configuration}
&\multicolumn{1}{c}{Term}
\\
\hline
\vspace{0.12mm}
 1  & 1s$^{2}$2s$^{2}$2p          & $^{2}$P       & 22 & 1s$^{2}$2s$^{2}$5g        & $^{2}$G   \\
 2  & 1s$^{2}$2s2p$^{2}$          & $^{4}$P       & 23 & 1s$^{2}$2s$^{2}$6s        & $^{2}$S   \\
 3  & 1s$^{2}$2s2p$^{2}$          & $^{2}$D       & 24 & 1s$^{2}$2s2p($^3$P$^0$)3p & $^{4}$D   \\
 4  & 1s$^{2}$2s2p$^{2}$          & $^{2}$S       & 25 & 1s$^{2}$2s2p($^3$P$^0$)3p & $^{2}$P   \\
 5  & 1s$^{2}$2s2p$^{2}$          & $^{2}$P       & 26 & 1s$^{2}$2s$^{2}$6p        & $^{2}$P$^{0}$   \\
 6  & 1s$^{2}$2s$^{2}$3s          & $^{2}$S       & 27 & 1s$^{2}$2s$^{2}$6d        & $^{2}$D   \\
 7  & 1s$^{2}$2s$^{2}$3p          & $^{2}$P$^{0}$ & 28 & 1s$^{2}$2s$^{2}$6f        & $^{2}$F$^{0}$   \\
 8  & 1s$^{2}$2p$^{3}$            & $^{4}$S$^{0}$ & 29 & 1s$^{2}$2s$^{2}$6g        & $^{2}$G   \\
 9  & 1s$^{2}$2s$^{2}$3d          & $^{2}$D       & 30 & 1s$^{2}$2s$^{2}$6h        & $^{2}$H$^{0}$   \\
 10 & 1s$^{2}$2p$^{3}$            & $^{2}$D$^{0}$ & 31 & 1s$^{2}$2s2p($^3$P$^0$)3p & $^{4}$S   \\
 11 & 1s$^{2}$2s$^{2}$4s          & $^{2}$S       & 32 & 1s$^{2}$2s$^{2}$7s        & $^{2}$S   \\
 12 & 1s$^{2}$2s$^{2}$4p          & $^{2}$P$^{0}$ & 33 & 1s$^{2}$2s($^3$P$^0$)3p   & $^{4}$P   \\
 13 & 1s$^{2}$2s2p($^3$P$^0$)3s   & $^{4}$P$^{0}$ & 34 & 1s$^{2}$2s$^{2}$7p        & $^{2}$P$^{0}$   \\
 14 & 1s$^{2}$2s$^{2}$4d          & $^{2}$D       & 35 & 1s$^{2}$2s$^{2}$7d        & $^{2}$D   \\
 15 & 1s$^{2}$2p$^{3}$            & $^{2}$P$^{0}$ & 36 & 1s$^{2}$2s$^{2}$7f        & $^{2}$F$^{0}$   \\
 16 & 1s$^{2}$2s$^{2}$4f          & $^{2}$F$^{0}$ & 37 & 1s$^{2}$2s$^{2}$7g        & $^{2}$G   \\
 17 & 1s$^{2}$2s$^{2}$5s          & $^{2}$S       & 38 & 1s$^{2}$2s$^{2}$7h        & $^{2}$H$^{0}$   \\
 18 & 1s$^{2}$2s$^{2}$5p          & $^{2}$P$^{0}$ & 39 & 1s$^{2}$2s2p($^3$P$^0$)3p & $^{2}$D   \\
 19 & 1s$^{2}$2s2p($^3$P$^{0})$3s & $^{2}$P$^{0}$ & 40 & 1s$^{2}$2s$^{2}$8g        & $^{2}$G   \\
 20 & 1s$^{2}$2s$^{2}$5d          & $^{2}$D       & 41 & 1s$^{2}$2s2p($^3$P$^0$)3d & $^{4}$F$^{0}$   \\
 21 & 1s$^{2}$2s$^{2}$5f          & $^{2}$F$^{0}$ &    & &    \\
\hline
\end{tabular}
\end{table}

\begin{table}
\tabcolsep1.7mm
\caption{As Table~\ref{levels_cii}, but for \CIII.}
\label{levels_ciii}
\begin{tabular}{rllrrc}
\hline
\hline
\multicolumn{1}{l}{C3\_\#}
&\multicolumn{1}{l}{Configuration}
&\multicolumn{1}{l}{Term}
&\multicolumn{1}{c}{C3\_\#}
&\multicolumn{1}{c}{Configuration}
&\multicolumn{1}{c}{Term}
\\
\hline
\vspace{0.09mm}
 1  & 1s$^{2}$2s$^{2}$        & $^{1}$S       & 36 & 1s$^{2}$2s5p            & $^{3}$P$^{0}$ \\
 2  & 1s$^{2}$2s2p            & $^{3}$P$^{0}$ & 37 & 1s$^{2}$2p($^2$P$^0$)3p & $^{1}$S \\
 3  & 1s$^{2}$2s2p            & $^{1}$P$^{0}$ & 38 & 1s$^{2}$2s5d            & $^{3}$D \\
 4  & 1s$^{2}$2p$^{2}$        & $^{3}$P       & 39 & 1s$^{2}$2s5g            & $^{3}$G \\
 5  & 1s$^{2}$2p$^{2}$        & $^{1}$D       & 40 & 1s$^{2}$2s5g            & $^{1}$G \\
 6  & 1s$^{2}$2p$^{2}$        & $^{1}$S       & 41 & 1s$^{2}$2s5d            & $^{1}$D \\
 7  & 1s$^{2}$2s3s            & $^{3}$S       & 42 & 1s$^{2}$2p($^2$P$^0$)3d & $^{1}$P$^{0}$ \\
 8  & 1s$^{2}$2s3s            & $^{1}$S       & 43 & 1s$^{2}$2s5f            & $^{3}$F$^{0}$ \\
 9  & 1s$^{2}$2s3p            & $^{1}$P$^{0}$ & 44 & 1s$^{2}$2s5f            & $^{1}$F$^{0}$ \\
 10 & 1s$^{2}$2s3p            & $^{3}$P$^{0}$ & 45 & 1s$^{2}$2s6s            & $^{3}$S \\
 11 & 1s$^{2}$2s3d            & $^{3}$D       & 46 & 1s$^{2}$2s6s            & $^{1}$S \\
 12 & 1s$^{2}$2s3d            & $^{1}$D       & 47 & 1s$^{2}$2s6p            & $^{3}$P$^{0}$ \\
 13 & 1s$^{2}$2p($^2$P$^0$)3s & $^{3}$P$^{0}$ & 48 & 1s$^{2}$2s6p            & $^{1}$P$^{0}$ \\
 14 & 1s$^{2}$2s4s            & $^{3}$S       & 49 & 1s$^{2}$2s6d            & $^{3}$D \\
 15 & 1s$^{2}$2p($^2$P$^0$)3s & $^{1}$P$^{0}$ & 50 & 1s$^{2}$2s6g            & $^{1}$G \\
 16 & 1s$^{2}$2s4s            & $^{1}$S       & 51 & 1s$^{2}$2s6g            & $^{3}$G \\
 17 & 1s$^{2}$2s4p            & $^{3}$P$^{0}$ & 52 & 1s$^{2}$2s6d            & $^{1}$D \\
 18 & 1s$^{2}$2p($^2$P$^0$)3p & $^{1}$P       & 53 & 1s$^{2}$2s6h            & $^{3}$H$^{0}$ \\
 19 & 1s$^{2}$2s4d            & $^{3}$D       & 54 & 1s$^{2}$2s6h            & $^{1}$H$^{0}$ \\
 20 & 1s$^{2}$2s4f            & $^{3}$F$^{0}$ & 55 & 1s$^{2}$2s6f            & $^{3}$F$^{0}$ \\
 21 & 1s$^{2}$2s4f            & $^{1}$F$^{0}$ & 56 & 1s$^{2}$2s6f            & $^{1}$F$^{0}$ \\
 22 & 1s$^{2}$2s4p            & $^{1}$P$^{0}$ & 57 & 1s$^{2}$2s7s            & $^{3}$S \\
 23 & 1s$^{2}$2p($^2$P$^0$)3p & $^{3}$D       & 58 & 1s$^{2}$2s7p            & $^{1}$P$^{0}$ \\
 24 & 1s$^{2}$2s4d            & $^{1}$D       & 59 & 1s$^{2}$2s7d            & $^{3}$D \\
 25 & 1s$^{2}$2p($^2$P$^0$)3p & $^{3}$S       & 60 & 1s$^{2}$2s7g            & $^{3}$G \\
 26 & 1s$^{2}$2p($^2$P$^0$)3p & $^{3}$P       & 61 & 1s$^{2}$2s7d            & $^{1}$D \\
 27 & 1s$^{2}$2p($^2$P$^0$)3d & $^{1}$D$^{0}$ & 62 & 1s$^{2}$2s7f            & $^{3}$F$^{0}$ \\
 28 & 1s$^{2}$2p($^2$P$^0$)3p & $^{1}$D       & 63 & 1s$^{2}$2s8p            & $^{1}$P$^{0}$ \\
 29 & 1s$^{2}$2p($^2$P$^0$)3d & $^{3}$F$^{0}$ & 64 & 1s$^{2}$2s8d            & $^{3}$D \\
 30 & 1s$^{2}$2p($^2$P$^0$)3d & $^{3}$D$^{0}$ & 65 & 1s$^{2}$2p9d            & $^{3}$D \\
 31 & 1s$^{2}$2s5s            & $^{1}$S       & 66 & 1s$^{2}$2p($^2$P$^0$)4s & $^{3}$P$^{0}$ \\
 32 & 1s$^{2}$2s5s            & $^{3}$S       & 67 & 1s$^{2}$2p($^2$P$^0$)4p & $^{1}$P \\
 33 & 1s$^{2}$2p($^2$P$^0$)3d & $^{3}$P$^{0}$ & 68 & 1s$^{2}$2p($^2$P$^0$)4p & $^{3}$D \\
 34 & 1s$^{2}$2p($^2$P$^0$)3d & $^{1}$F$^{0}$ & 69 & 1s$^{2}$2p($^2$P$^0$)4p & $^{3}$P \\
 35 & 1s$^{2}$2s5p            & $^{1}$P$^{0}$ & 70 & 1s$^{2}$2p($^2$P$^0$)4p & $^{1}$D \\
\hline                     
\end{tabular}
\end{table}

\begin{table}
\tabcolsep1.7mm
\caption{As Table~\ref{levels_cii}, but for \CIV.}
\label{levels_civ}
\begin{tabular}{rllrrc}
\hline
\hline
\multicolumn{1}{l}{C4\_\#}
&\multicolumn{1}{l}{Configuration}
&\multicolumn{1}{l}{Term}
&\multicolumn{1}{c}{C4\_\#}
&\multicolumn{1}{c}{Configuration}
&\multicolumn{1}{c}{Term}
\\
\hline
\vspace{0.09mm}
 1  & 1s$^{2}$2s & $^{2}$S       & 26 & 1s$^{2}$7i  & $^{2}$I    \\
 2  & 1s$^{2}$2p & $^{2}$P$^{0}$ & 27 & 1s$^{2}$7h  & $^{2}$H$^{0}$   \\
 3  & 1s$^{2}$3s & $^{2}$S       & 28 & 1s$^{2}$8s  & $^{2}$S   \\
 4  & 1s$^{2}$3p & $^{2}$P$^{0}$ & 29 & 1s$^{2}$8p  & $^{2}$P$^{0}$   \\
 5  & 1s$^{2}$3d & $^{2}$D       & 30 & 1s$^{2}$8d  & $^{2}$D   \\
 6  & 1s$^{2}$4s & $^{2}$S       & 31 & 1s$^{2}$8f  & $^{2}$F$^{0}$   \\
 7  & 1s$^{2}$4p & $^{2}$P$^{0}$ & 32 & 1s$^{2}$8g  & $^{2}$G   \\
 8  & 1s$^{2}$4d & $^{2}$D       & 33 & 1s$^{2}$8h  & $^{2}$H$^{0}$   \\
 9  & 1s$^{2}$4f & $^{2}$F$^{0}$ & 34 & 1s$^{2}$8i  & $^{2}$I   \\
 10 & 1s$^{2}$5s & $^{2}$S       & 35 & 1s$^{2}$9s  & $^{2}$S   \\
 11 & 1s$^{2}$5p & $^{2}$P$^{0}$ & 36 & 1s$^{2}$9p  & $^{2}$P$^{0}$   \\
 12 & 1s$^{2}$5d & $^{2}$D       & 37 & 1s$^{2}$9d  & $^{2}$D   \\
 13 & 1s$^{2}$5f & $^{2}$F$^{0}$ & 38 & 1s$^{2}$9f  & $^{2}$F$^{0}$   \\
 14 & 1s$^{2}$5g & $^{2}$G       & 39 & 1s$^{2}$9g  & $^{2}$G   \\
 15 & 1s$^{2}$6s & $^{2}$S       & 40 & 1s$^{2}$9h  & $^{2}$H$^{0}$   \\
 16 & 1s$^{2}$6p & $^{2}$P$^{0}$ & 41 & 1s$^{2}$9i  & $^{2}$I   \\
 17 & 1s$^{2}$6d & $^{2}$D       & 42 & 1s$^{2}$10p & $^{2}$P$^{0}$   \\
 18 & 1s$^{2}$6f & $^{2}$F$^{0}$ & 43 & 1s$^{2}$10d & $^{2}$D   \\
 19 & 1s$^{2}$6g & $^{2}$G       & 44 & 1s$^{2}$11p & $^{2}$P$^{0}$   \\
 20 & 1s$^{2}$6h & $^{2}$H$^{0}$ & 45 & 1s$^{2}$11d & $^{2}$D   \\
 21 & 1s$^{2}$7s & $^{2}$S       & 46 & 1s$^{2}$12p & $^{2}$P$^{0}$   \\
 22 & 1s$^{2}$7p & $^{2}$P$^{0}$ & 47 & 1s$^{2}$12d & $^{2}$D   \\
 23 & 1s$^{2}$7d & $^{2}$D       & 48 & 1s$^{2}$13p & $^{2}$P$^{0}$   \\
 24 & 1s$^{2}$7f & $^{2}$F$^{0}$ & 49 & 1s$^{2}$13d & $^{2}$D   \\
 25 & 1s$^{2}$7g & $^{2}$G       & 50 & 1s$^{2}$14d & $^{2}$D   \\
\hline
\end{tabular}
\end{table}

\section{Dependence on stellar parameters} 
\label{app2}

\begin{figure*}
\center
{\includegraphics[scale=0.75,angle=90]{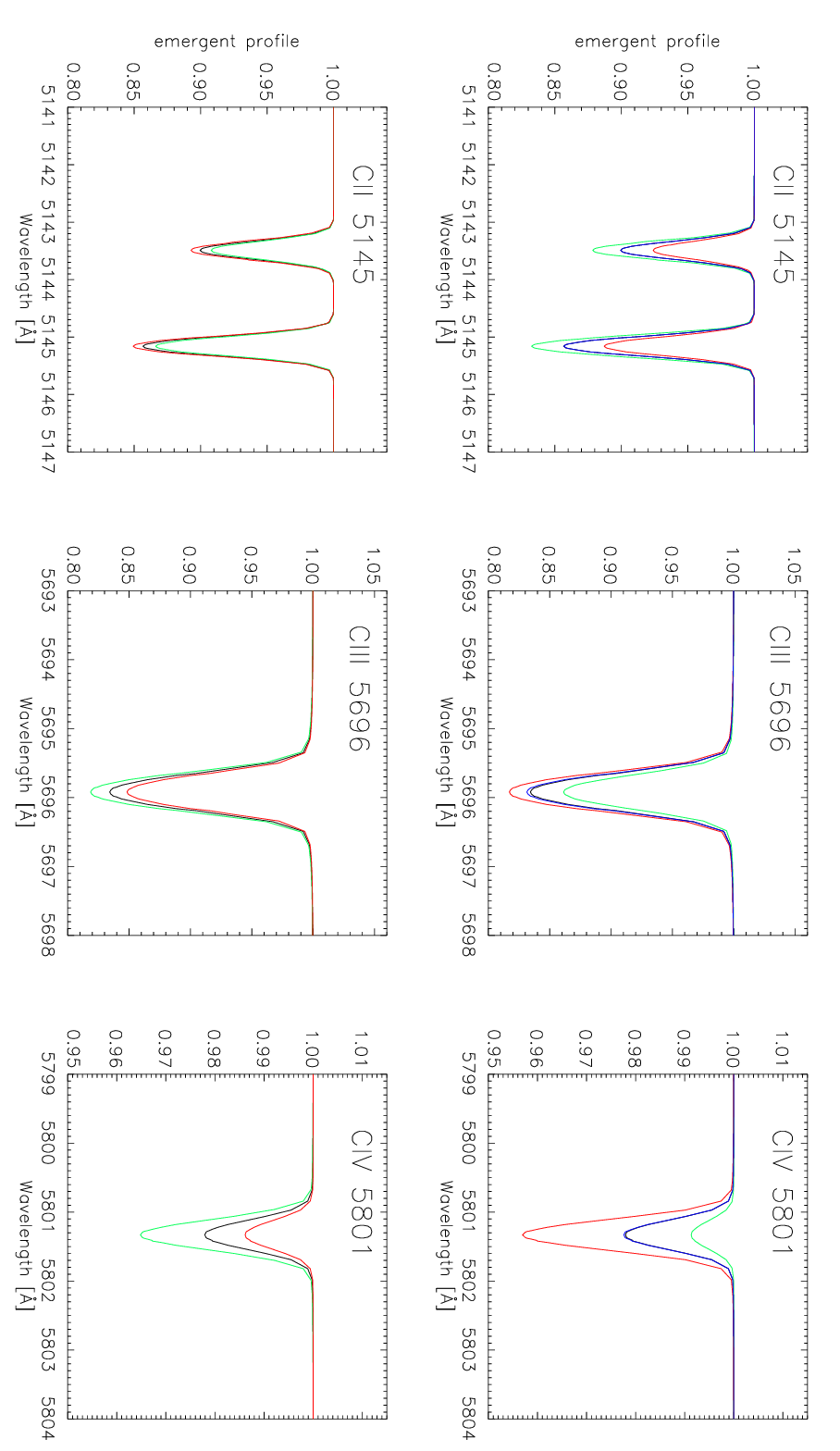}}
%
\caption{\CII\,5145, \CIII\,5696, and \CIV\,5801 line profiles
for model D30 (black lines, see Table~\ref{tab_grid}) and similar
models with relatively small changes in effective temperature (\Teff)
and gravity (\logg). In the upper panels, the red lines correspond to
a D30 model with \Teff\ increased by 1000~K, the green lines to a
model with \Teff\ decreased by the same value, while the blue lines
display the reaction to a decrease of \mdot\ by a factor of three. In
the lower panels, the red and green lines correspond to a D30 model with \logg\
increased and decreased by 0.15~dex, respectively}
\label{d30_cprof}
\end{figure*}

\begin{figure*}
\center
{\includegraphics[scale=0.75,angle=90]{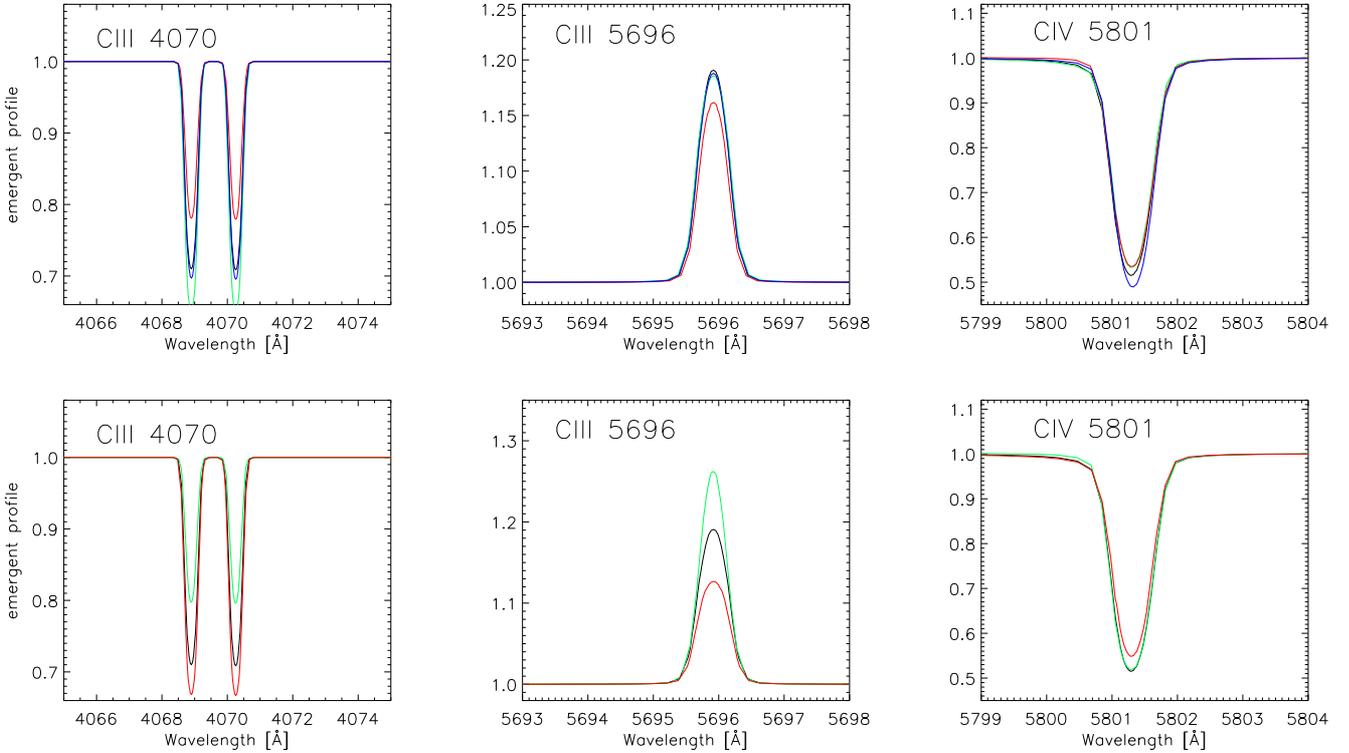}}
\caption{As Fig.~\ref{d35_cprof}, but for model D40, and \CIII\,4068-70,
instead of \CII\,5145.}
\label{d40_cprof}
\end{figure*}

\begin{figure*}
\center
{\includegraphics[scale=0.75,angle=90]{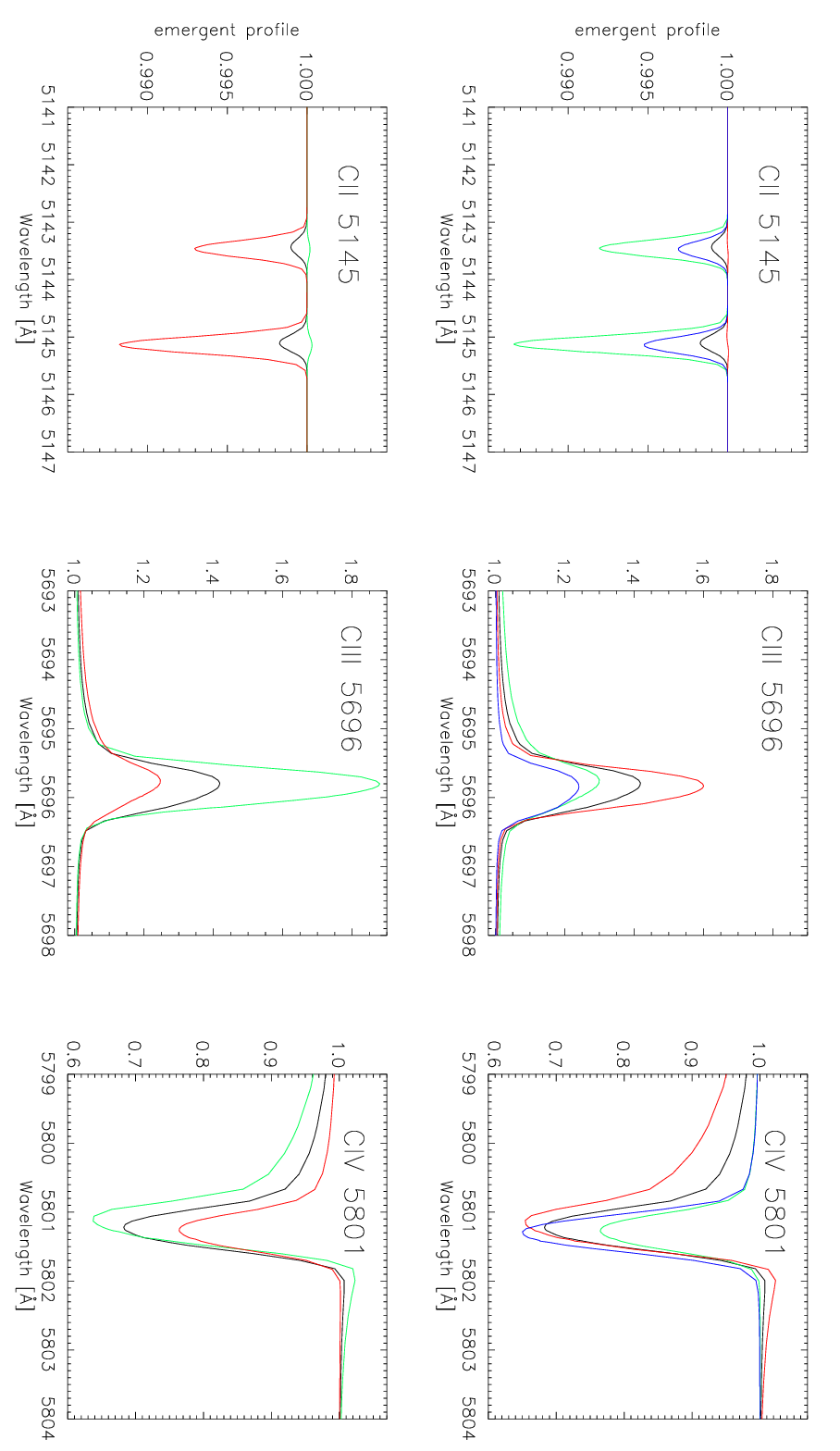}}
\caption{As Fig.~\ref{d30_cprof}, but for model S30, and $\Delta \logg = 0.1$.}
\label{s30_cprof}
\end{figure*}

\begin{figure*}
\center
{\includegraphics[scale=0.75,angle=90]{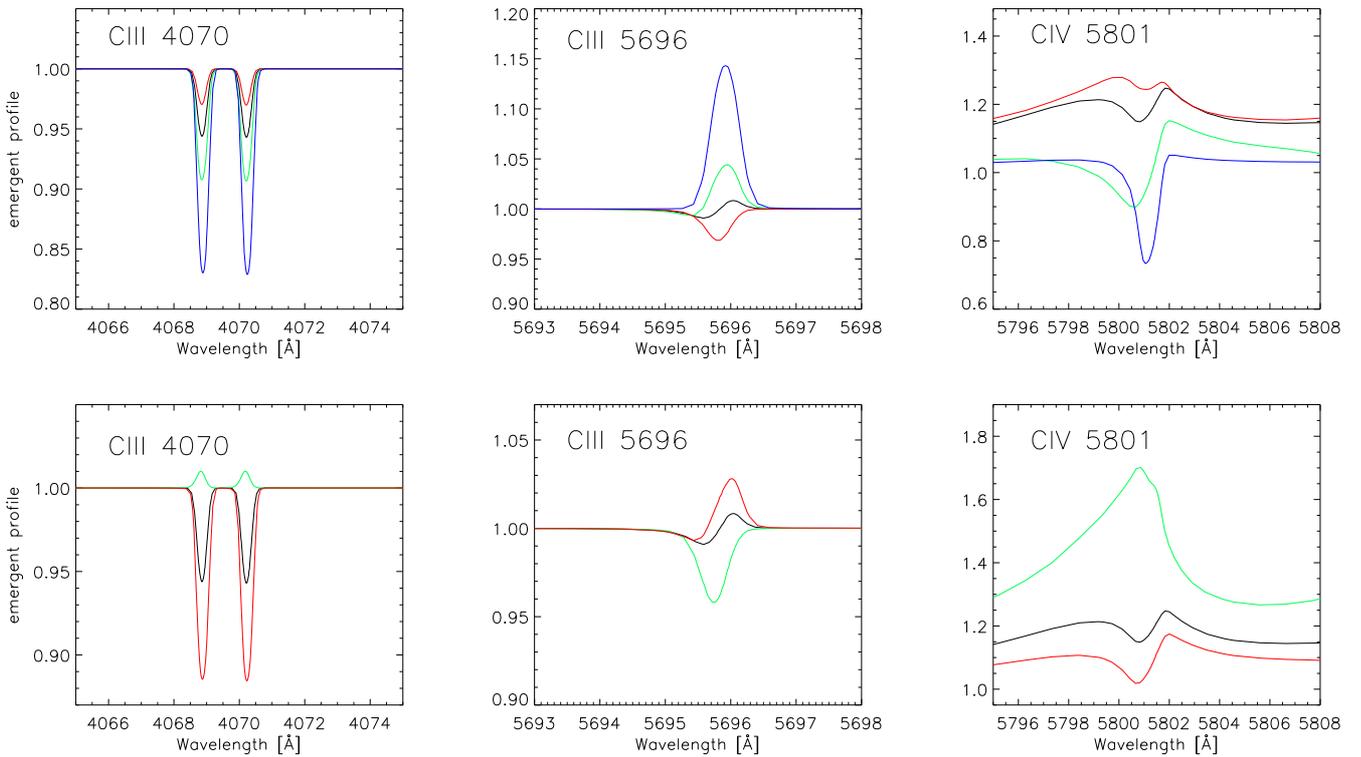}}
\caption{As Fig.~\ref{d40_cprof}, but for model S40.}
\label{s40_cprof}
\end{figure*}

This appendix displays the sensitivity of synthetic carbon spectra
from dwarf and supergiant models at 30 and 40~kK, with respect to
variations in \Teff, \logg, and \mdot.
Figs.~\ref{d30_cprof} to \ref{s40_cprof} have the same 
layout as Fig.~\ref{d35_cprof}, and refer to Sect.~\ref{para_dep}. 

\clearpage

\end{document}